\documentclass[times, twocolumn]{aastex631_arxiv}
\usepackage{CJK}
\usepackage{graphicx}
\usepackage{subfigure}
\usepackage{enumerate}
\usepackage{amssymb, amsmath}
\usepackage{natbib}
\usepackage{color}
\usepackage{ulem}
\usepackage{dblfnote}
\usepackage{appendix}
\usepackage{hyperref}
\usepackage[stable]{footmisc}
\usepackage{threeparttable}
\usepackage{realboxes}

\usepackage[nodayofweek]{datetime}
\newdateformat{monthyearday}{%
  \THEYEAR\ \monthname[\THEMONTH] \twodigit{\THEDAY}}
  
\submitjournal{AAS Journals}
\shorttitle{JWST/NIRCam Coronagraphic Imaging of HD~163296}
\shortauthors{Uyama et al.}

\begin{document}

\title{JWST/NIRCam Coronagraphic Search for Hidden Planets in the HD~163296 Protoplanetary Disk}


\correspondingauthor{Luca Ricci}
\email{luca.ricci@csun.edu}

\author[0000-0002-6879-3030]{Taichi Uyama}
    \affiliation{Department of Physics and Astronomy, California State University Northridge, 18111 Nordhoff Street, Northridge, CA 91330, USA}
    \affiliation{Astrobiology Center, 2-21-1 Osawa, Mitaka, Tokyo 181-8588, Japan}
\author[0000-0001-8123-2943]{Luca Ricci}
    \affiliation{Department of Physics and Astronomy, California State University Northridge, 18111 Nordhoff Street, Northridge, CA 91330, USA}
\author[0000-0001-7591-2731]{Marie Ygouf}
    \affiliation{Jet Propulsion Laboratory, California Institute of Technology, 4800 Oak Grove Dr., Pasadena, CA, 91109, USA}
\author[0000-0003-2253-2270]{Sean Andrews}
    \affiliation{Center for Astrophysics, Harvard \& Smithsonian, 60 Garden St., Cambridge, MA 02138, USA}  
\author{Sara Gallagher}
    \affiliation{Department of Physics and Astronomy, California State University Northridge, 18111 Nordhoff Street, Northridge, CA 91330, USA}
\author[0000-0001-6947-6072]{Jane Huang}
    \affiliation{Department of Astronomy, Columbia University, 538 W. 120th Street, Pupin Hall, New York, NY 10027, USA}
\author[0000-0001-8061-2207]{Andrea Isella}
    \affiliation{Department of Physics and Astronomy, Rice University, 6100 Main St, Houston, TX 77005, USA}
    \affiliation{Rice Space Institute, Rice University, 6100 Main St, Houston, TX 77005, USA}
\author[0000-0002-8895-4735]{Dimitri Mawet}
    \affiliation{Department of Astronomy, California Institute of Technology, 1200 E. California Blvd., Pasadena, CA 91125, USA}
    \affiliation{Jet Propulsion Laboratory, California Institute of Technology, 4800 Oak Grove Dr., Pasadena, CA, 91109, USA}
\author[0000-0002-1199-9564]{Laura P{\'e}rez}
    \affiliation{Departamento de Astronomía, Universidad de Chile, Camino El Observatorio 1515, Las Condes, Santiago, Chile}
\author[0000-0002-9573-3199]{Massimo Robberto}
    \affiliation{Space Telescope Science Institute, 3700 San Martin Drive, Baltimore, MD 21218, USA}
    \affiliation{Department of Physics \& Astronomy, Johns Hopkins University, 3400 N. Charles Street, Baltimore, MD 21218, USA}
\author[0000-0003-4769-1665]{Garreth Ruane}
    \affiliation{Jet Propulsion Laboratory, California Institute of Technology, 4800 Oak Grove Dr., Pasadena, CA, 91109, USA}
\author[0000-0002-8537-9114]{Shangjia Zhang}
    \altaffiliation{NASA Hubble Fellowship Program Sagan Fellow}
    \affiliation{Department of Astronomy, Columbia University, 538 W. 120th Street, Pupin Hall, New York, NY 10027, USA}
    \affiliation{Department of Physics and Astronomy, University of Nevada, Las Vegas, 4505 S. Maryland Pkwy, Las Vegas, NV, 89154, USA}
    \affiliation{Nevada Center for Astrophysics, University of Nevada, Las Vegas, Las Vegas, NV 89154, USA}
\author[0000-0003-3616-6822]{Zhaohuan Zhu}
    \affiliation{Department of Physics and Astronomy, University of Nevada, Las Vegas, 4505 S. Maryland Pkwy, Las Vegas, NV, 89154, USA}
    \affiliation{Nevada Center for Astrophysics, University of Nevada, Las Vegas, Las Vegas, NV 89154, USA}

\begin{abstract}

HD~163296 is a Herbig Ae/Be star with multiple signposts of on-going planet formation on its disk, such as prominent rings and gaps, as well as kinematic features as identified by previous ALMA observations. We carried out JWST/NIRCam coronagraphic imaging using the F410M and F200W NIRCam filters, with the goal of detecting the emission from the putative young planets in this system. Our F410M observations did not detect the putative planets at the predicted locations of the ALMA velocity kinks, but detected a point-like source candidate at a separation of $\approx0\farcs75$ and a position angle of $\approx231\fdg4$ that is unlikely a background star because of the measured flux in the F410M filter and the detection limit in the F200W filter. These data achieved unprecedented contrast levels at $\sim4~\micron$ at stellocentric separations $\rho\gtrsim0\farcs8$. This allowed us to derive stringent constraints at the outer velocity kink ($\Delta {\rm F410M}=15.2~{\rm mag}$) on the mass of the putative planet with or without a circumplanetary disk, and considering different possible initial entropies for the planet.

\end{abstract}


\section{Introduction} 
\label{sec: Introduction}

The discovery of thousands of exoplanets over the last couple of decades has shown that the birth of planets is a very efficient process in nature \citep[e.g.,][]{Burke2015}. However, several of the physical mechanisms responsible for their formation and evolution are still poorly understood. Starting with the very first detection of an exoplanet around a Solar-like star \citep{Mayor1995}, the discovery of hot Jupiters very close to their star has suggested that the orbits of planets can significantly vary during their life \citep{Lin1996}.
A natural explanation for this migration of planets involves the gravitational interaction between the still-forming planet and the parental disk \citep[see, e.g., the review by][]{Baruteau2014}. The theory of planet-disk interaction predicts that planets interact with the disk by launching density waves, and the consequent torque on the planet can lead to planet migration.

Although theoretical investigations of the disk-planet interaction have been conducted for several decades \citep{Lin1979,Goldreich1979}, this theory still needs empirical validation. Observations at optical and near infrared wavelengths predominantly trace stellar radiation scattered by fine dust grains on the upper layers of the disk, and may not directly reflect the spatial distribution of solids in the planet-forming disk regions. The distribution of solids in the dense disk regions can be investigated at longer sub-mm/mm wavelengths, for example with ALMA, because of the lower optical depths of the dust emission.

Altogether, these observations of nearby protoplanetary disks have unveiled annular gaps, cavities, and spiral arms in the disk that may be produced by the interaction with planets \citep[for recent reviews, see][]{Bae2023,Paardekooper2023}. 
The most common disk substructures are concentric gaps and rings observed in the sub-mm continuum with ALMA \citep[e.g.,][]{Andrews2018,Huang2018}. In several cases the morphology of these structures are consistent with the predictions of models of planet-disk interaction, and their observational properties have been used to estimate the mass of the putative planets \citep[e.g.,][]{Kanagawa2015,Dipierro2018,Zhang2018,Lodato2019}.

Although concentric rings and gaps, as well as most of the other observed structures, are compatible with the hypothesis of one or more embedded planets in those systems, they are not unique to that scenario. In fact, several processes, such as molecular condensation fronts \citep{Banzatti2015}, photoevaporation \citep{Owen2010}, and also various magneto-hydrodynamical instabilities have been shown to produce comparable structures in the disk - e.g., magneto-rotational instability \citep{Flock2015}; zonal flows \citep{Uribe2011}; radially variable magnetic disk winds \citep{Suriano2018}, and vertical shear instability \citep{Flock2015}. Given the multi-dimensional parameter space of the models, the best way to discriminate between scenarios of disk-planet interaction vs disk chemical/physical mechanisms is to detect planets.

A particularly interesting disk where to study planet formation and planet-disk interaction is the HD~163296 system. 
HD~163296 is a young intermediate mass \citep[$M_{\star} \approx 2~M_{\odot}$,][]{Andrews2018} Herbig Ae star located at $100.97\pm0.42$~pc from Earth \citep{Gaia2016,Gaia2023}. The star is surrounded by a gaseous disk with a radius of about 500~au. In contrast, mm-wave continuum emission arising from solid particles is confined within about 200 au from the star \citep{Isella2007,deGregorio-Monsalvo2013}. \cite{Isella2018} and \cite{Huang2018} imaged this disk using ALMA in the dust continuum at 1.3 mm with an angular resolution of about 0\farcs04, or 4~au at the distance of the source. These observations unveiled four annular concentric gaps (or dark annuli) centered at about 0\farcs10, 0\farcs48, 0\farcs86, and 1\farcs45 from the star, corresponding to stellocentric radii of 10 (gap D10), 48 (D48), 86 (D86), and 145~au (D145), respectively. The analysis of the radial profile of the continuum surface brightness shows that the bright rings are radially narrow, and interleaved by wide gaps. For example, the inferred radial widths of the D48 and D86 gaps are of about $20-30$~au, and for the D145 gap of about $40-50$~au, respectively \citep{Isella2016,Isella2018}. These are among the widest gaps detected within the disks imaged at high-resolution with ALMA by the Disk Substructures at High Angular Resolution Project \cite[DSHARP;][]{Andrews2018}.

The outer gaps D86 and D145 show depletion also in CO gas density \citep{Isella2016}. The widths and levels of mass depletion inferred for the dust and gas density of the HD~163296 disk favor the hypothesis of young giant planets opening those gaps, and are less consistent with alternative scenarios involving chemical processes such as molecular condensation fronts. Furthermore, the high mass accretion rate $\dot{M} \approx 5 \times 10^{-7}~M_{\odot}$ yr$^{-1}$ derived for HD~163296 \citep{Mendigutia2013} is less compatible also with disk photoevaporation from high-energy stellar photons to form any of these gaps.
\cite{Liu2018} used the \texttt{LA-COMPASS} hydrodynamic code \citep{Li2005,Li2009} to reproduce the multi-ring structure seen in HD~163296. This code accounts for the gravitational torques induced by planets in the disk, as well as the aerodynamic interaction between gas and dust. The D86 and D145 gaps are consistent with being opened by young giant planets with masses close to the one of Jupiter. Similar results were obtained, and extended to the inner gaps D10 and D48, by \cite{Zhang2018}.

Moreover, ALMA observations of the CO molecular line emission have detected significant deviations from the expected emission for a disk in Keplerian rotation, which are consistent with perturbations induced by the gravitational interaction with Jupiter-mass planets in the D86 and D145 outer gaps \citep{Teague2018,Izquierdo2022,Izquierdo2023}. \cite{Teague2019} analyzed the 3D kinematical structure of the disk and found evidence for meridional (vertical) flows of gas that they reproduced using disk models containing planets with masses of $\approx 1-2~M_{\rm{Jup}}$ in the D86 and D145 gaps, as well as at an outer orbit of 237 au from the central star. More localized perturbations in the kinematics of the Keplerian disk, known as 'velocity kinks', have been reported by \cite{Pinte2018,Pinte2020} in the ALMA channel maps for the $^{12}$CO ($J = 2-1$) emission line at angular separations from the star and position angles (East of North) of (2\farcs20, -$3^\circ$) for kink~\#1 and (0\farcs67, $93^\circ$) for \#2, respectively.
These locations are consistent with the D86 gap and with the stellocentric radius of the outer meridional flow \citep{Teague2019}.  Also these localized kinematical features are reproduced by models of disk-planet interaction with planets with a mass of $\approx 2~M_{\rm{Jup}}$.

Young disks with wide and cleared annular gaps are therefore excellent targets to attempt the direct detection of young planets and to test the predictions of planet-disk interaction models.
However, detecting newborn planets is challenging: the variability of young stars and the presence of dusty disks hinder the use of radial velocities and transit techniques. Direct imaging techniques at optical-infrared wavelengths have the potential to detect young planets that are distant from the star ($>5$~au) and that have cleared the tenuous external disk regions to expose themselves, though the number of detections is small \citep[see a review by][]{Currie2023}. With the exception of the convincing embedded planets in the PDS~70 system \citep[][]{Keppler2018,Haffert2019} and some other planet candidates \citep[e.g., AB~Aur~b, MWC~758c, and HD~169142b;][]{Currie2022,Wagner2023,Hammond2023}, other high-contrast imaging surveys did not detect planets embedded in disks \citep[e.g.,][]{Uyama2017,Cugno2019,Asensio-Torres2021,Cugno2023,Wallack2024}.
Several high-contrast imaging observations searched for embedded planets also in the HD~163296 system \citep[e.g., ][]{Guidi2018,Mesa2019,Rich2019,Xie2020,Hasegawa2024}, but did not detect any convincing planet signal. 
We carried out new coronagraphic observations of the HD~163296 system using the \textit{Near Infrared Camera} (NIRCam) aboard the \textit{James Webb Space Telescope} (JWST). 
The main goal of these observations is to investigate the presence of the planets that have been proposed to explain the disk structures observed in this system.

In this paper, we present the results of the new JWST/NIRCam observations of the HD~163296 system. Section~\ref{sec: Data} presents the main characteristics of these observations together with the methods adopted for the data reduction; the results of these observations are outlined in Section~\ref{sec: Results} and discussed in Section~\ref{sec: Discussion}. Section~\ref{sec: Conclusion} summarizes the main findings of this work.

\section{Observations and Data Reduction} \label{sec: Data}
\subsection{NIRCam Observations}

We observed HD~163296 with JWST/NIRCam \citep{Rieke2005} under the Cycle~1 GO~2540 program (PI: Luca Ricci) \footnote{All the data used in this paper can be found in MAST: \dataset[https://doi.org/10.17909/j87q-jf82]{https://doi.org/10.17909/j87q-jf82}.}.

Primary observations of the HD~163296 system utilized the Lyot round masks in two filters, i.e. MASK210R for F200W in the SHORT channel (SUB640A210R subarray, pixel scale~$\approx 0\farcs031/{\rm pix}$), and MASK430R for F410M in the LONG channel (SUB320A430R subarray, pixel scale~$\approx 0\farcs063/{\rm pix}$), respectively. NIRCam enables simultaneous exposures at the SHORT and LONG channels while one Lyot mask is used. 
In total, we obtained four-configuration observations: primary F200W/MASK210R and F410M/MASK430R, and by-product F200W/MASK430R and F410M/MASK210R.
Our program did not observe a PSF reference star for reference-star differential imaging (RDI) because HD~163296 is known to have a strong infrared excess emission due to circumstellar dust very close to the star, but employed roll-subtraction angular differential imaging (ADI) by obtaining the data at two roll positions for post-processing to remove the stellar PSF. 
The readout pattern was set to 'MEDIUM8', and the total exposure times were 9220.69~seconds per roll (19841.38~seconds in total) for the F410M/MASK430R and F200W/MASK430R observations, and 8057.75~seconds per roll (16115.48~seconds in total) for the F200W/MASK210R and F410M/MASK210R observations. The roll angle difference for ADI was $10\fdg35$.

The secondary by-product configurations, i.e. F200W/MASK430R and F410M/MASK210R, however, were not as useful as the primary ones for the detection of embedded planets: in the F200W/MASK430R image the central star is located close to the edge of the field of view (FOV), and the F410M/MASK210R image suffers from intense starlight residuals within a few arcseconds from the central star, leaving significantly more residual speckles and producing worse contrast levels than in the primary data with MASK430R (see Appendix \ref{sec: Comparison of the two masks at F410M}).
Hereafter, `F200W' and `F410M' refer to the primary configurations of F200W/MASK210R and F410M/MASK430R, respectively.

\subsection{NIRCam Data Reduction} \label{sec: Data Reduction}
\subsubsection{Pre-processing} \label{sec: Pre-processing}


We downloaded the stage~0 datasets ({\tt \verb|*|uncal.fits}), and then used the spaceKLIP pipeline\footnote{See the \href{https://github.com/kammerje/spaceKLIP}{spaceKLIP documentation} for detailed explanations and a tutorial of the pipeline.} \citep[version 2.1.1;][]{Kammerer2022,Carter2023} for data calibration \citep[with the {\tt JWST} pipeline;][for stage~0 to 2]{Bushouse2023}, which is customized for coronagraphic imaging \citep[see][]{Carter2023}. The version of JWST calibration reference data system (CRDS) file we used is 11.17.14.

Particularly before the post-processing steps to search for faint sources around HD~163296 by PSF subtraction (stage~2 to 3), we read the stage~2 datasets ({\tt \verb|*|calints.fits}) into the spaceKLIP {\tt Database} class and then operated pre-processing functions in the {\tt imageTools} class to conduct background subtraction, bad-pixel correction (masking and interpolation), image centering (from the mask center to the PSF centroid) and alignment (correcting frame-to-frame shifts). To further mitigate the effect of potential speckles that are not static during the observations, we applied pixel-wise sigma-clipping (3$\sigma$) for the integrations in each \texttt{FITS} file after the image alignment process. We also excluded outlier integration frames that deviate by $>3\sigma$ from the averaged integration image and those with noisy speckles, which we identified by visually checking all integration frames. 
In total we removed one integration frame (corresponding to an exposure time of about 62 seconds) from the F410M data set and nine frames (exposure time of about 263.7 seconds) from the F200W data set.

\subsubsection{Post-processing} \label{sec: Post-processing}
After pre-processing, we aimed at subtracting stellar halo leaking the coronagraph mask and instrumental speckles that hinder faint planet signals from the science exposures by producing the best representation of the PSF. 
We made use of the {\tt pyKLIP} modules in the {\tt imageTools} class, which is a statistical post-processing pipeline used for differential imaging data through the Karhunen-Lo\`eve Image Processing algorithm \citep[KLIP;][]{Soummer2012}.
Regarding the {\tt pyKLIP} parameters, we did not divide the FoV into several subsections with the `annuli' and `subsections' parameters because there are numerous objects (e.g., background objects and the disk features, see Section \ref{sec: Results}) in the FoV, and dividing the FoV into several subsections could induce artifacts, particularly in the presence of bright nearby sources. Instead, we cropped the FoV into a $10\arcsec\times 10\arcsec$ area before running {\tt pyKLIP} modules.  
As previous studies predicted embedded protoplanets in the HD~163296 disk at stellocentric separations $\rho\lesssim2\farcs5$, this paper focuses on sources within $2\farcs5$ from the central star, while outer sources will be discussed in a future work.
As the field rotation for ADI is $\approx10$~deg, the most aggressive PSF subtraction can attenuate faint or extended features; in order to mitigate this effect, we adopted small values for KL modes for moderately aggressive ADI reduction when showing our post-processed images in Section~\ref{sec: Results}, although we also show the post-processing results which can be obtained with other values for the KL modes in Appendix~\ref{sec: Gallery of the post-processed images with different pyKLIP parameters}.

As mentioned in Section~\ref{sec: Pre-processing} we did not observe a PSF reference star. Moreover, at the moment there are no archival data for reference stars observed in the same configurations as our program. We attempted to employ RDI by generating synthesized reference PSFs using WebbPSF \citep{Perrin2014}, and incorporating wavefront errors taken on the closest date to our observations. However, we did not achieve as high-contrast performance as with the ADI reduction (see Appendix \ref{sec: Attempt at RDI with a synthesized WebbPSF} for the results of this attempt using RDI).

\section{Results} \label{sec: Results}

\begin{figure*}
    \centering
        \includegraphics[width=\linewidth]{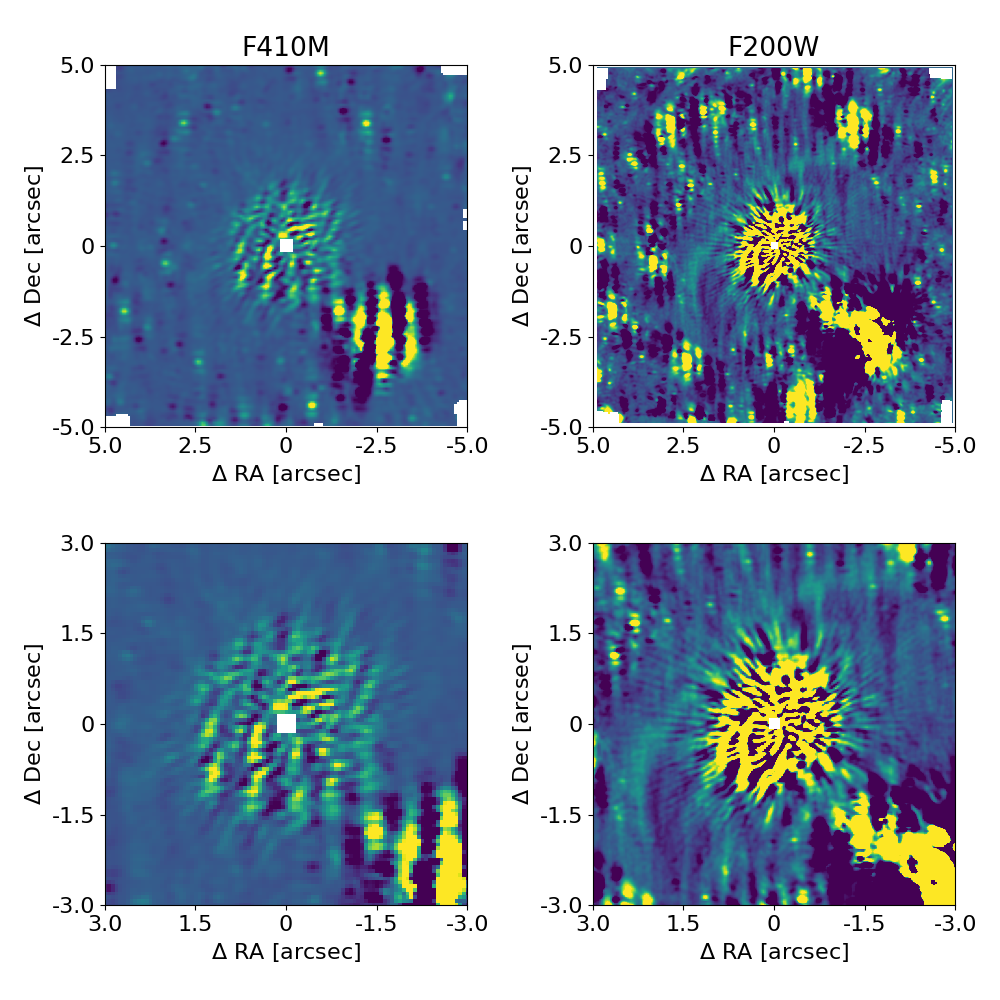}
        \caption{Top: Post-processed JWST/NIRCam maps in the F410M (left) and F200W (right) filters, obtained with KL=3. Bottom: Zoomed-in images of the post-processed results (top).}
        \label{fig: NIRCam post-processed}
\end{figure*}

\begin{figure*}
    \centering
    \includegraphics[width=\textwidth]{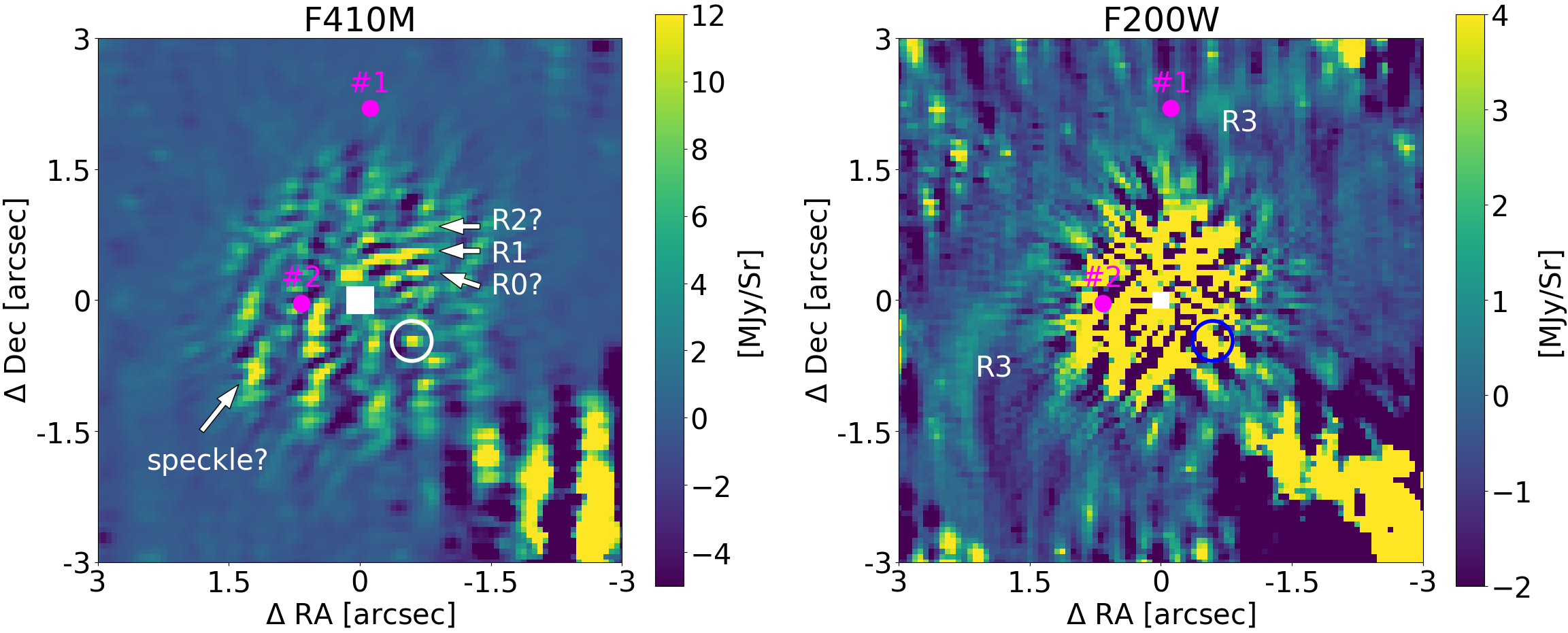}
    \caption{NIRCam F410M (left) and F200W (right) maps as in the bottom panels in Figure~\ref{fig: NIRCam post-processed} with symbols representing the predicted locations of putative protoplanets from the velocity kinks observed with ALMA \citep[indicated by magenta dots with the kink ids;][]{Pinte2020}, a point-like source candidate (white and blue circles in the left and right panels, respectively), and ring features (R0--R3).}
    \label{fig: F200W vs F410M}
\end{figure*}

\subsection{Post-processed results}

The post-processed JWST/NIRCam images are shown in Figure~\ref{fig: NIRCam post-processed}.
We detected numerous sources in both filters but in this study we focus on the 6\arcsec\ by 6\arcsec\ region around the central star (bottom panels in Fig.~~\ref{fig: NIRCam post-processed}) where we expect planets in the HD~163296 disk. 
Especially in our F200W map, a difference is evident in the number density of background sources between the regions within and beyond the HD~163296 outermost ring feature (at separations $\rho\sim2\farcs5-3\farcs5$, see below); this is a clear signature of the extinction of light from background sources due to dust in the disk \citep[see also theoretical predictions in][]{Sanchis2020,Marleau2022,Hasegawa2024,Alarcon2024}. We note that the elongated features visible towards the South-West corner in the F200W map is a cluster of bright background sources with diffraction patterns from the NIRCam PSF and self-subtraction features caused by the ADI reduction \citep[cf. SPHERE HD~163296 data;][]{Mesa2019,Juillard2024}.

Our NIRCam observations did not detect significant IR emission from the putative protoplanets at the locations of the velocity kinks observed with ALMA, but at the same time some interesting features are visible on the images: 1) a compact source with a contrast of $1.8\times10^{-5}$ at $\rho\approx0\farcs75$ and a position angle of $\approx231\fdg4$ in the F410M image, which does not have a counterpart in the F200W image (see Figure~\ref{fig: F200W vs F410M}); 2) two arc-like features were clearly detected, such as R1 ($\rho\sim0\farcs45$ to the North of the star) in the F410M image and R3 ($\rho\sim2\farcs2$ to the North) in the F200W image (see Figure~\ref{fig: F200W vs F410M}), while two other similar features are less clear, such as R0 (very close to the star) and R2 ($\rho\sim0\farcs7$ to the North); in the rest of the paper we refer to these features as \textit{rings} because these arc-like features are likely the forward-scattering side of more extended ring structures, as shown by previous studies \citep[e.g.,][]{Guidi2018,Mesa2019,Rich2019,Juillard2024};
3) the resolved ring feature in the F200W image (R3) is in close proximity with the outer velocity kink (see the right panel in Figure~\ref{fig: F200W vs F410M}). Detailed discussions of these features are provided in the following subsections.

\subsection{Search for Embedded Protoplanets} \label{sec: Search Protoplanets}

\begin{figure*}
    \centering
    \includegraphics[width=0.9\textwidth]{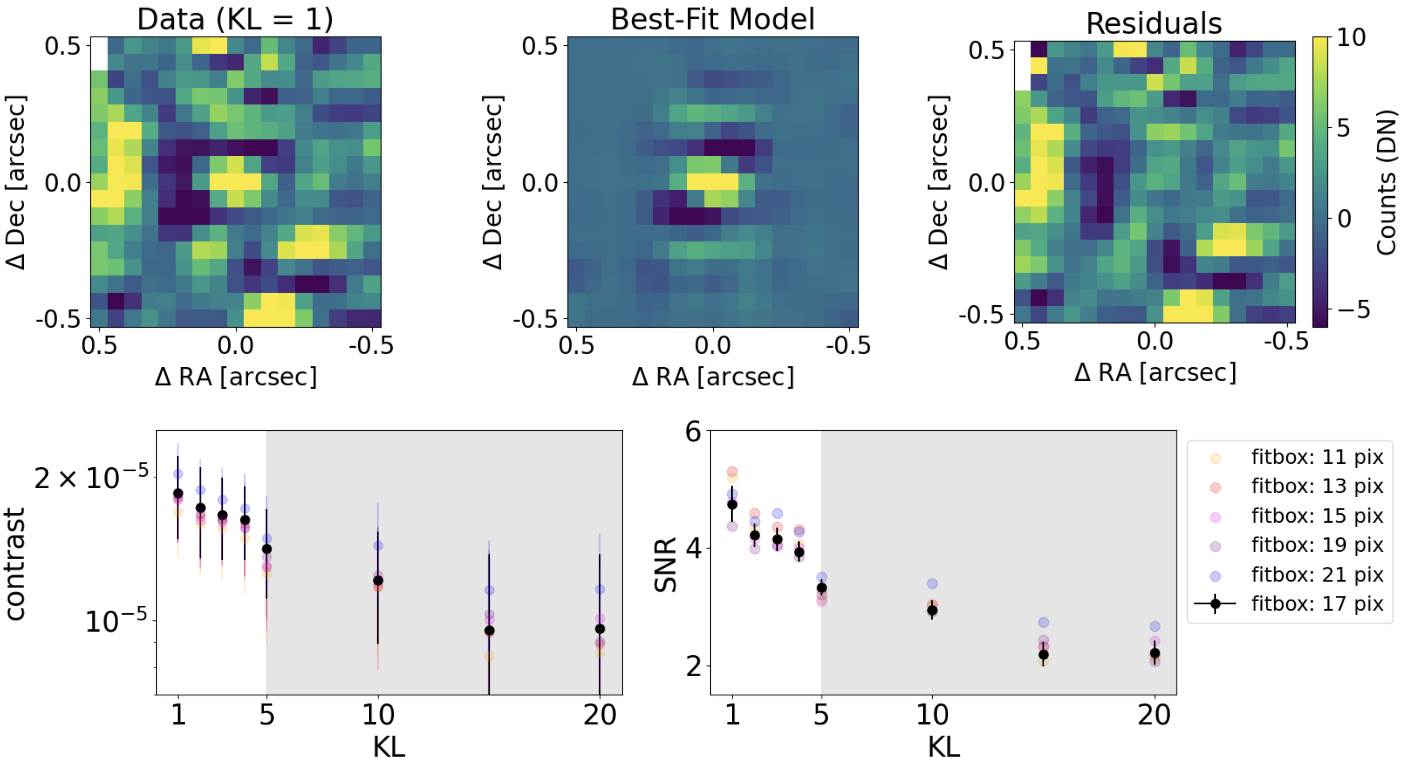}
    \caption{PSF fitting results of the source candidate with KL=1 (top panels), and the derived contrast and SNR values for different KL modes (bottom panels). The colors in the plot indicate the box size used for the PSF fitting, demonstrating that the extracted photometry and SNR do not largely depend on the size parameter. We finally adopted the 17-pix box corresponding to $\sim1\arcsec \times 1\arcsec$ (see text). The error bars of the 17-pix plot in the right panel correspond to the standard deviations of the SNR measurements among the 13--21 pixel range. The gray-shaded parameter space (${\rm KL}\geq5$) indicates aggressive PSF subtraction and a heavy self-subtraction effect onto astronomical sources at $\sim0\farcs75$, where the source candidate is located. See also Appendix~\ref{sec: Gallery of the post-processed images with different pyKLIP parameters} for the PSF fitting results at different KL modes.
    }
    \label{fig: PSF fitting}
\end{figure*}

We searched for point-like sources on the post-processed F410M image within the disk region. In order to consider a point-like feature as a source candidate we requested that 1) the point-like source is surrounded by adjacent self-subtraction features caused by ADI reduction, and that 2) its signal does not significantly vary at different KL modes. Note that large KL modes lead to aggressive PSF subtraction causing heavy self-subtraction, leading to attenuation of any signal particularly at small separations (see also Appendix~\ref{sec: Gallery of the post-processed images with different pyKLIP parameters}), and we prioritized small KL modes. 
We then attempted forward modeling and PSF fitting using the {\tt extract\_companions} module \citep[cf, {\tt pyklip.fitpsf}; ][]{pyklip} for quantitative investigations. 
We found one point-like feature (indicated by a circle in Figure~\ref{fig: F200W vs F410M}) is well fitted with the forward-modeled PSF. 
Table~\ref{tab: compact sources} summarizes the photometry and relative astrometry and Figure \ref{fig: PSF fitting} shows the PSF fitting result at different KL modes. The noise here was measured as the standard deviation on the residual map within a local $\sim1\arcsec\times1\arcsec$ box around the source candidate (the upper right panel in Figure~\ref{fig: PSF fitting}). Typically the noise in high-contrast imaging is measured as the standard deviation within annular regions assuming the noise to be azimuthally symmetric. However, our data show the disk feature that can contribute to noise and azimuthally asymmetric speckle residuals due to the small field-rotation ADI processing and the PSF of NIRCam coronagraphic imaging. 
We note that the SNR and the extracted contrast decrease at KL$\geq5$ because of a significant self-subtraction effect via aggressive ADI subtraction at such a small separation from the star, making the extracted photometry and SNRs at such large KLs unreliable (cf, Appendix~\ref{sec: Gallery of the post-processed images with different pyKLIP parameters}). In contrast at small KLs (KL = 1--4) the extracted contrast level is more stable.
There are several other relatively bright features at marginal SNRs, but they are spatially extended, suggesting that they are likely residual speckles (e.g. the annotated extended feature ('speckle?') in the left panel of Figure~\ref{fig: F200W vs F410M}).

\begin{table}[]
\begin{center}
    \caption{Information on the source candidate from the F410 NIRCam map (KL~=~1).}
    \begin{tabular}{cccc}
      Stellocentric Separation & Position Angle & Contrast & SNR \\ \hline\hline   
      $0\farcs754 \pm 0\farcs007$ & $231\fdg39\pm1\fdg27$ & $1.83\times10^{-5}$ & 5.0
    \end{tabular}
    \label{tab: compact sources}
\end{center}
\end{table}

PSF fitting of the compact source at KL=1 provides a contrast estimate of $1.8\times10^{-5}$, corresponding to an apparent magnitude in F410M of $15.03\pm0.15$.
Regarding the flux of the central star, which was used to convert the contrast estimate into an apparent magnitude, we collected available photometric information from VO Sed Analyzer \citep[VOSA;][]{Bayo2008} and read it into spaceKLIP, which derived the corresponding stellar magnitude at each filter, i.e. $4.97\pm0.02$~mag at F200W, and $3.20\pm0.07$~mag at F410M. 
Given the relatively marginal SNR of this source, future observations will be needed to confirm this detection. In Section~\ref{sec: Point-like Source}, we discuss the plausibility of different scenarios for the nature of this source given the information at our disposal.

\begin{figure*}
    \centering
    \includegraphics[width=\textwidth]{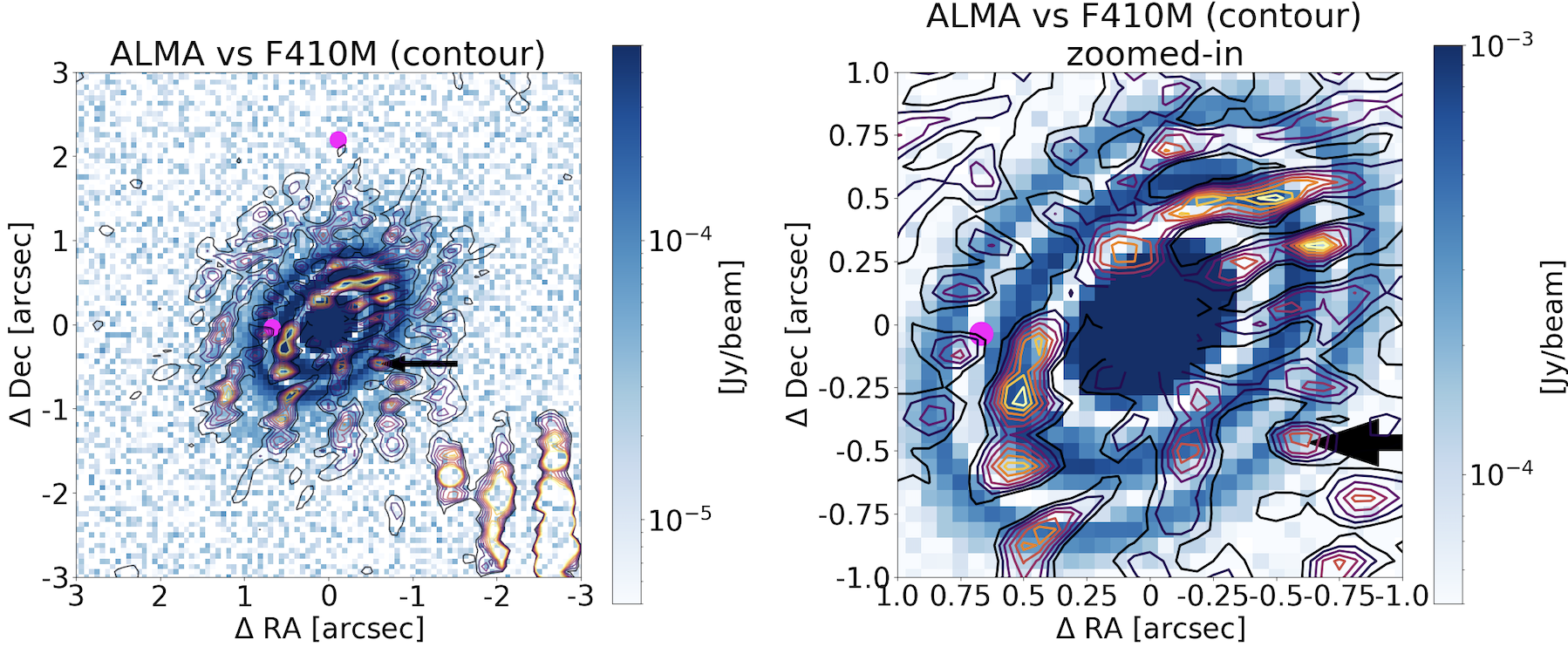}
    \caption{Comparison between the F410M map (contours) and the DSHARP ALMA continuum map (colors) at $\lambda = 1.3$ mm \citep[$6\arcsec\times6\arcsec$ map in the left and $2\arcsec\times2\arcsec$ map in the right, respectively;][]{Isella2018}. The location of the source candidate is indicated by a black arrow. The magenta dots show the location of the ALMA velocity kinks. }
    \label{fig: F410M vs ALMA}
\end{figure*}

\subsection{Disk Features} \label{sec: Disk Features}

\begin{figure}
\centering
    \includegraphics[width=0.45\textwidth]{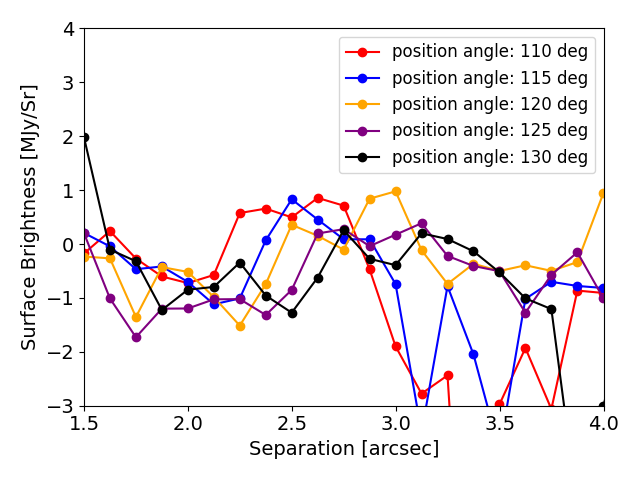}
    \caption{Radial profiles of the surface brightness at different position angles in the F200W image. The bump feature between $\sim2\arcsec-3\arcsec$ at each position angle indicates the location of R3. The outer regions beyond R3 contain signal from background stars as well as negative self-subtraction features, making the radial profiles noisier than in the region inside of R3.}
    \label{fig: surface brightness F200W}
\end{figure}

Our NIRCam observations in the F410M filter resolved the ring feature R1, which was reported by previous ground-based high-contrast imaging observations \citep[e.g.,][]{Guidi2018,Mesa2019,Rich2019,Juillard2024}. 
Although the R1 ring is also marginally seen in the F200W post-processed image, regions at such inner separations ($\rho\lesssim1\arcsec$) are speckle-dominant. Other ground-based high-contrast observations have achieved higher contrast levels within 1" from the star at NIR wavelengths and detected the R1 feature more clearly \citep[e.g.,][]{Mesa2019,Rich2019}.

The F410M map shows potential additional ring-like features other than R1 (see the left panel of Figure \ref{fig: F410M vs ALMA} and Appendix~\ref{sec: Gallery of the post-processed images with different pyKLIP parameters} for different KL modes). However, the regions in the map with separations from the star $\rho\lesssim1\farcs5$ are affected by residual speckles and the bright disk feature R1 might induce other diffraction patterns, which can mimic arc-like features.
Tentative evidence for R2 was found in an archival Keck/NIRC2 $Ms$-band dataset that marginally detected another arc-like feature just outside of R1 (see Appendix~\ref{sec: NIRC2}). R2 is seemingly collocated with an extended feature right outside of R1 in the F410M post-processed image (denoted as 'R2?' in Figure~\ref{fig: F200W vs F410M}). The presence of R2 suggests that the companion candidate reported in \cite{Guidi2018} may be part of R2.

To test whether these arc-like features in the F410M data are systematic diffraction patterns induced by R1 or real features (R1+R2), we forward-modeled a toy disk model with properties that resemble the R1 ring. As detailed in Appendix~\ref{sec: Forward modeling of the disk features}, we found that this ring did not induce additional significant extended features. 

We note that R2 has been detected only at $\lambda\gtrsim 3~\mu{\rm m}$ with Keck/NIRC2 and JWST/NIRCam while our F200W data as well as with previous high-contrast imaging at NIR wavelengths do not show evidence of R2 \citep[e.g.,][]{Rich2019,Juillard2024}. 
As mentioned in \cite{Guidi2018}, R1 is not co-located with any of the ALMA bright rings, most likely because scattered-light and dust thermal emission originate from different areas in a disk structure \citep[cf.,][]{Ginski2016,Dong2018,Boehler2018}. On the semi-minor axis direction \citep[at a position angle of $\approx43^\circ$;][]{Huang2018}, R1 has an offset of $\sim0\farcs2$ from the spine of B67 and the whole feature overlaps with the ALMA rings/gaps (D45, B67, and D86; see the right panel in Figure~\ref{fig: F410M vs ALMA}), suggesting that R1 is generated by scattering of the light from the central star by a flared inner wall of B67 \citep[cf., B67 is known to have a significantly large scale height;][]{Doi2021}. Similarly, R2 may also be due to scattered light at a flared inner wall of B100.

Although at low SNR, the F410M post-processed image also shows a marginal arc-like feature that is located inside of R1 (labeled as 'R0?' in Figure~\ref{fig: F200W vs F410M}), which appears to be collocated with a potential extended feature reported by previous Keck/NIRC2 $L'$-band observations \citep[labeled as 's' in][]{Guidi2018}. Follow-up observations with higher contrast are needed to confirm the R0 and R2 features, and help to shed light on the morphology of the disk surface.

The F200W image resolves another scattering ring feature in the outermost disk regions (R3, semi-major axis~$\sim3\farcs5$, width: $\sim0\farcs5$; see the middle panel in Figure~\ref{fig: F200W vs F410M}). Tentative evidence for this structure was presented from radial cuts of the disk surface brightness by HST coronagraphic imaging in the optical \citep{Grady2000,Wisniewski2008}. \cite{Rich2020} revisited this archival data set and better resolved the ring feature. Our NIRCam observations clearly confirmed this R3 feature.
Figure~\ref{fig: surface brightness F200W} illustrates the radial profiles of the surface brightness around R3 at position angles of 110--130~degrees, showing a $\sim0\farcs5$-wide with a peak at $\sim1-1.2$~MJy/Sr (note that these radial profiles were not corrected for self-subtraction induced by the ADI reduction).
\cite{Muro-Arena2018} modeled the dust profile using detection limits from VLT/SPHERE polarimetric imaging and our result will be useful to further constrain the dust characteristics of the HD~163296 disk.

\cite{Pinte2018} argued that the velocity kink~\#1 observed with ALMA is collocated with the gap detected with HST in the outer disk \citep[corresponding to R3 in Figure~\ref{fig: F200W vs F410M};][]{Grady2000,Rich2020}, but our JWST/NIRCam map suggests that the kink feature is collocated with the R3 ring itself rather than with the gap. However, as mentioned above, the disk surface is likely flared and the disk regions around R3 could also be flared. 
Future detailed modeling of the disk-planet interaction will shed light on how the putative planet at kink~\#1 may interact with the disk and produce features consistent with the observations presented in this work.

\subsection{Contrast Limits} \label{sec: Contrast Limits}

We calculated the contrast limits of the JWST/NIRCam observations using the {\tt AnalysisTools} modules after masking bright background sources. Negative self-subtraction features are not fully masked, particularly for those at larger separations from the central star and the disk features are not masked either, which increases the noise component and affects the contrast curves. 

\begin{figure*}
\centering
\begin{tabular}{cc}
\begin{minipage}{0.5\hsize}
    \centering
    \includegraphics[width=0.95\textwidth]{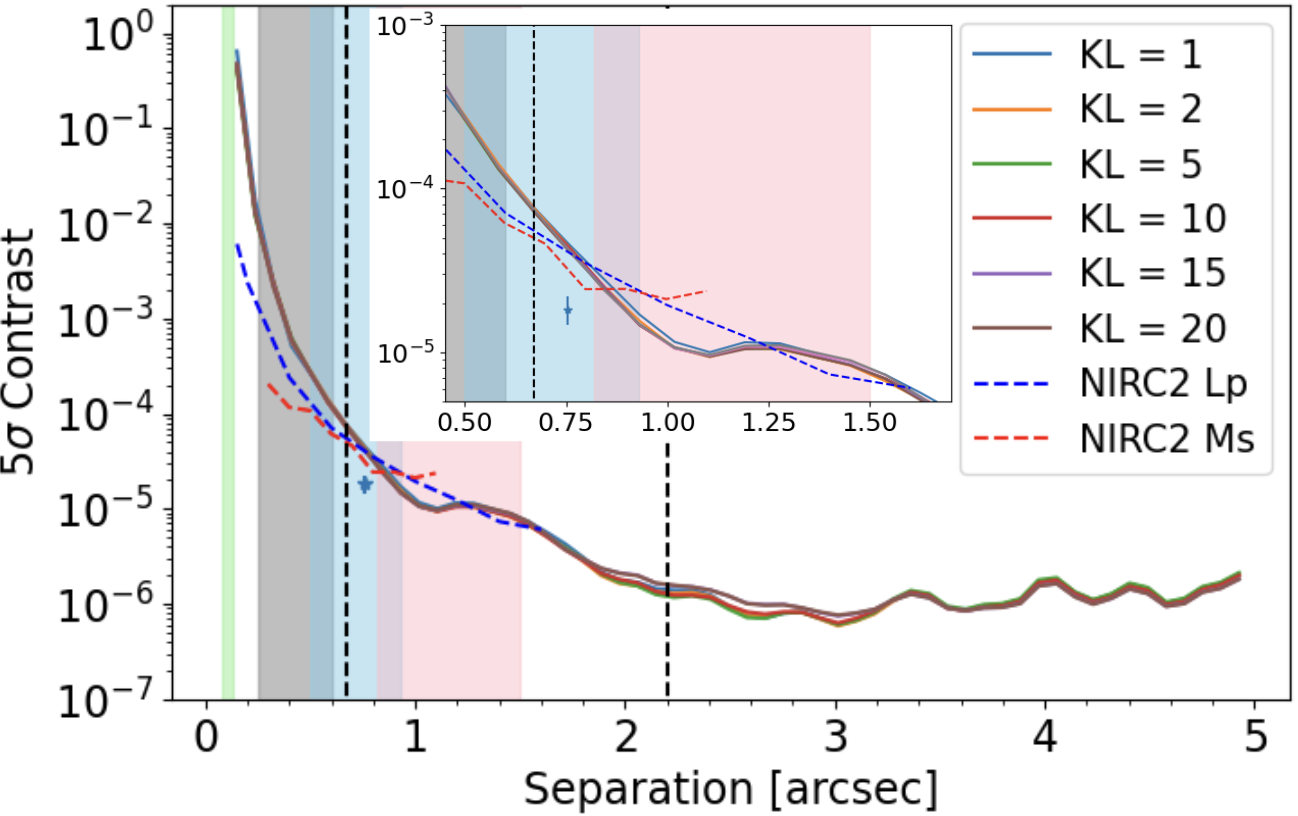}
\end{minipage}
\begin{minipage}{0.5\hsize}
    \centering
    \includegraphics[width=\textwidth]{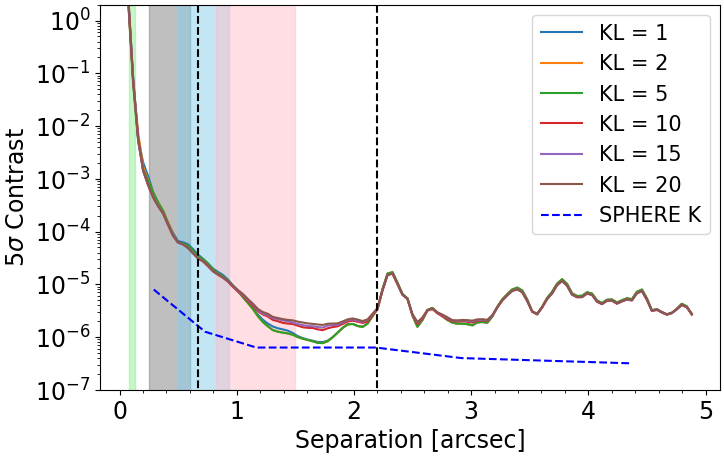}
\end{minipage}
\end{tabular}
\caption{Left) 5$\sigma$ contrast limits of the F410M data overlaid with Keck/NIRC2 contrast limits \citep[$L'$-band;][$Ms$-band; this study, see Appendix~\ref{sec: NIRC2}]{Wallack2024}. The insert panel is a zoomed-in version of the contrast limits at separations from the central star around 1\arcsec. The star symbol indicates the contrast and separation of the source candidate from the F410M data. The vertical dashed black lines indicate the separations of the ALMA velocity kinks. The light-green, gray, sky-blue, and pink shaded areas correspond to the projected separation ranges of the ALMA dust gaps \citep[D10, D45, D86, and D145, respectively;][]{Huang2018,Isella2018}. Right) The contrast limits obtained from the F200W data overlaid with a rough SPHERE 5$\sigma$ contrast limit we inspected visually from \cite{Mesa2019}, where we applied the disk inclination to the separations for the consistency with our contrast limits.}
\label{fig: contrast limit}
\end{figure*}

The azimuthally-averaged radial profiles of the F410M 5$\sigma$ contrast limits are presented in the left panel of Figure~\ref{fig: contrast limit}, overlaid with the detection limits from the NIRC2 $L'$- and $Ms$-band deep coronagraphic observations of HD~163296 \cite[the $L'$-band observations are from][and the $Ms$-band observations are presented in Appendix~\ref{sec: NIRC2}]{Wallack2024}. As mentioned in Section~\ref{sec: Search Protoplanets}, this method calculates azimuthally-averaged noise values at a given range for the separation from the central star, and this is different from the SNR estimate of the PSF fitting. Particularly at the speckle-dominant separations ($\rho\lesssim1\farcs5$) where the residual speckles are not fully azimuthally symmetric, the azimuthally-averaged contrast limit may have overestimated the noise level at a specific position. Therefore we could fit the source candidate that is fainter than the typical F410M contrast limit at $\sim0\farcs75$ (star symbol in the left panel of Figure~\ref{fig: contrast limit}).
Note that these contrast curves were calculated without deprojecting the disk as the 'noise' is attributed by the residual starlight and speckles that are irrelevant to the disk geometry.
Our observations achieved a deeper contrast limit than NIRC2 at $\rho\gtrsim0\farcs8$ (note that the observational configurations, such as filters, coronagraph specifications, total exposure time, and field rotations, are different). 

We also calculated the contrast limits of the F200W data in the same way as the F410M data (see the right panel of Figure~\ref{fig: contrast limit}). Note that the F200W data contain numerous background stars with bright diffraction patterns as well as negative self-subtraction features (see Figure~\ref{fig: NIRCam post-processed}) and it is technically hard to mask all of them when calculating the standard deviation for the noise estimate, making it noisier than the background sensitivity of NIRCam/F200W even outside the regions dominated by speckle noise ($\rho\gtrsim1\farcs5$).
Previous ground-based high-contrast imaging in the NIR with VLT/SPHERE and Subaru/CHARIS achieved better contrast limits than the F200W data within 1\arcsec\ \citep[e.g.,][]{Mesa2019,Rich2019}.

The predicted masses of the putative protoplanets in the HD 163296 system are $\approx1-2\ M_{\rm Jup}$ or smaller \citep[e.g.,][]{Liu2018,Pinte2018,Pinte2020,Teague2019} and evolutionary models for these objects \citep{Spiegel2012} predict significantly brighter fluxes at $\sim$ 4\micron\ than at 2\micron.  Given the contrast limits derived from our NIRCam observations, the F410M observations are more sensitive to the emission of these planets than the F200W ones. We therefore used the F410M results to estimate upper limits to putative embedded protoplanets and also considered possible circumplanetary disk emission in Section~\ref{sec: Constraining the Mass of the ALMA Putative Planets}.

\section{Discussion} \label{sec: Discussion}

As outlined in Section~\ref{sec: Results}, our JWST/NIRCam observations detected a compact source candidate at $\approx 0\farcs75$ from the HD~163296 star in the F410M filter, multiple rings in scattered light in the F200W and F410M filters, and achieved unprecedented contrast limits that allow us to derive upper limits to the mass of putative protoplanets in this system. In this section we discuss in more details the implications of these results.

\subsection{Nature of the Point-like Source Candidate} \label{sec: Point-like Source}

\begin{figure}
    \centering
    \includegraphics[width=0.47\textwidth]{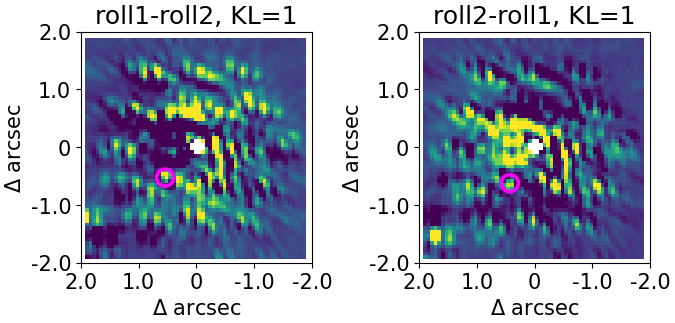}
    \caption{Comparison of roll-subtracted images at each roll at KL=1. The magenta circle indicates the location of the source candidate. Note that the north position angle of each image is not corrected.}
    \label{fig: roll-sub image}
\end{figure}

The source candidate detected in F410M (Section~\ref{sec: Search Protoplanets}) does not have a counterpart in the F200W map. 
The contrast limit from the F200W image is $\sim2\times10^{-5}$, corresponding to an apparent magnitude of $\sim$16.7, and the lower limit of the F200W - F410M color is 1.7. 
Furthermore, previous ground-based observations achieving better contrast levels than the F200W contrast limit (see Section~\ref{sec: Contrast Limits}) did not report any counterpart to the F410W source candidate. \cite{Mesa2019} achieved a contrast limit of $\sim10^{-6.2}$ in $K$-band, corresponding to apparent magnitude of $\sim20.3$, and the lower limit of the $K$ - F410M color of this source is $\sim5.3$, suggesting a very red color.

Assuming that this source candidate is a real object, in order to derive the intrinsic magnitude of a source from the measured magnitude one needs to know the extinction in that filter. This is critical for protoplanets that are still embedded in their parent disk, as it has been shown that the dust in the disk can cause very high extinction values even at mid-IR wavelengths for a planet in the midplane which has not opened a significant gap along its orbit \citep[e.g.,][]{Sanchis2020,Alarcon2024}. The extinction rapidly decreases for planets which are massive enough to clear out the disk material close to their orbit. However, given the large uncertainties on the mass estimates of (putative) young planets, and the intrinsic difficulty in estimating dust opacities and column densities throughout the disk, an estimate of the local extinction from the disk at the location of the putative planet is at this point highly uncertain.

HD~163296 does not show any significant extinction~\citep[$A_{\rm V}=0$;][]{Rich2019}. If an embedded planet has cleared out the disk around its orbit so that light from the planet is not extincted by the surrounding disk materials, our measured magnitude at F410M (see Fig.~\ref{fig: mass limit} with $\Delta$F410M~=~11.8~mag) would correspond to a hot-start $\approx2-4\ M_{\rm Jup}$ giant planet \citep{Spiegel2012}. These planet mass estimates would be lowered if part of the measured flux originates from a circumplanetary disk (CPD) surrounding the putative planet. For example, as detailed in Section~\ref{Circumplanetary}, according to the CPD models by \cite{Zhu2018}, the measured contrast level would correspond to a warm/cold-start accreting planet with the product between planet mass and mass accretion rate of $\approx 6\times10^{-6}M_{\rm Jup}^2/{\rm yr}$, with a CPD with an inner radius of $\approx 1~R_{\rm Jup}$.

We note that, depending on the disk properties (e.g., viscosity) massive planets would likely perturb the gas and dust in the disk \citep[$>10\%$ of Keplerian motion, corresponding to $\gtrsim2\ M_{\rm Jup}$ planets at a few tens of au;][]{Pinte2020}. Under the assumption that this source is a real planet embedded in the midplane of the disk, the deprojected stellocentric separation corresponds to $\sim111~{\rm au}$, just outside B100 (see Figure~\ref{fig: F410M vs ALMA}). The separation of the source candidate coincides with a dip in the radial profile of the $^{12}$CO emissions (Figure~\ref{fig: radial profile of CO, MAPS}, see also Appendix~\ref{sec: Companion candidate vs ALMA CO velocity maps} for the comparison with the CO map), which is likely due to a local decrease in the gas density of the disk.
A future detailed analysis of the disk-planet interaction using hydrodynamic numerical models can shed more light on the possible perturbations induced by this object on the disk, as well as on the expected extinction due to the disk material at the location of the source candidate. 

\begin{figure*}
    \centering
    \includegraphics[width=0.8\textwidth]{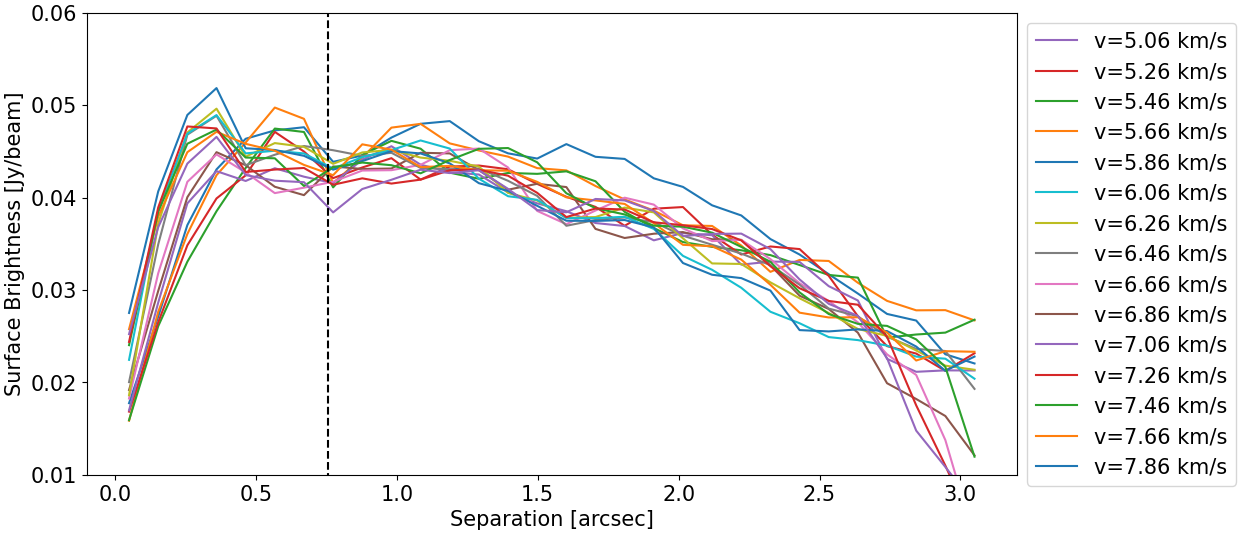}
    \caption{Radial profiles of the South West side of the $^{12}$CO gas emissions at $v=6.46\pm1.4~{\rm km /s}$ (see Appendix~\ref{sec: Companion candidate vs ALMA CO velocity maps}). The vertical dashed line indicates the separation of the source candidate.}
    \label{fig: radial profile of CO, MAPS}
\end{figure*}

In another scenario in which no local perturbation to the disk is present, \cite{Alarcon2024} derived a very high extinction value of $A_{\rm F410M} \sim 25-30$~mag at the location of our source candidate (see also Section~\ref{sec: Constraining the Mass of the ALMA Putative Planets} for the model assumptions and caveats). 
If true, this would make the intrinsic luminosity at 4~\micron\ of this source even higher than the central star, and it would make very unlikely also the scenario of a background object for the nature of this source.

Even though the source candidate is detected with a robust SNR at small KL modes, however, it is still possible that this feature is actually false positive after averaging. Due to the NIRCam diffraction patterns, extended feature like R1 could introduce a point-like feature at different locations, but our forward modeling test with the toy model presented in Appendix~\ref{sec: Forward modeling of the disk features} did not reproduce such a bright and stable feature at the location of the source candidate. Therefore we do not consider this scenario hereafter.

We also investigated intermediate files of roll-subtracted images at each roll state to investigate the possibility of a residual speckle or a hot/bad-pixel cluster. If there was a bright feature at either of the roll-subtracted images, it suggests that the feature is likely a variable speckle or a hot/bad pixel cluster. If such feature appears at both roll-subtracted images, it suggests real astronomical source. Figure~\ref{fig: roll-sub image} shows the combined roll-subtracted images and we did not recognize significant excesses suggesting such false positives at the location of the source candidate we detected. However, we also did not confirm the stable signal there because each of the roll-subtracted state images has noisier speckle residuals than the final processed image. 

In summary, the extracted color and location relative to the HD~163296 disk suggests that this source candidate is intriguing if it is a real astronomical source, but because of technical difficulties of our data, we defer the conclusion of this identity to follow-up observations with a higher contrast or future analysis with RDI reduction when reference PSFs in the same observational configurations are available.

\subsection{Constraining the Mass of the ALMA Putative Planets} \label{sec: Constraining the Mass of the ALMA Putative Planets}

\subsubsection{Comparison with Evolutionary Models}

\begin{figure}
    \centering   
    \includegraphics[width=0.45\textwidth]{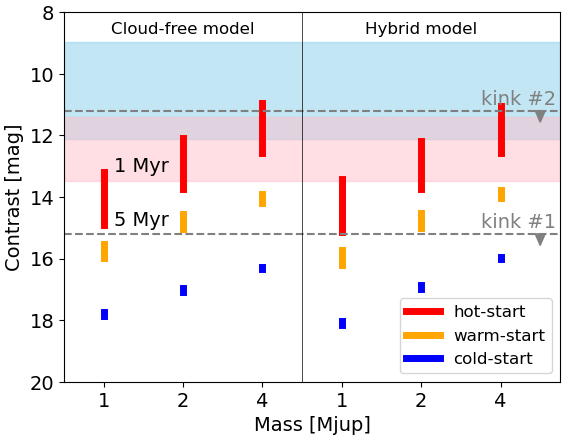}
\caption{Diagram showing the conversion between the contrast limits in the F410M filter and the corresponding predicted contrast from evolutionary models of young giant planets with different properties. The dashed horizontal lines indicate the F410M empirical detection limits at the location of the ALMA velocity kink \#1 ($\Delta$F410M=15.2~mag) and \#2 ($\Delta$F410M=11.2~mag), respectively. The colored shades indicate the contrast limits at the projected separations of the ALMA dust gaps (sky-blue for D45 and pink for D86, as in Figure~\ref{fig: contrast limit}). We took into account hot- (red)/warm- (orange)/cold-start (blue) evolutionary models \citep{Spiegel2012} with cloud-free and hybrid clouds atmospheres at solar metallicity \citep[][see text for details]{Burrows2011}, ages between 1 and 5~Myr (the upper and lower limits of each model range correspond to 1 and 5~Myr, respectively), and assuming the same extinction as the central star, i.e. $A_{\rm V}=0$.}
\label{fig: mass limit}
\end{figure}

We derived upper limits for the mass of the putative planets at the locations of the ALMA velocity kinks \citep[\#1 at $\rho=2\farcs2$ and \#2 at 0\farcs67;][]{Pinte2020} and dust gaps, with the exception of the D10 gap (light-green in Figure~\ref{fig: contrast limit}), which is located inside the inner working angle of the NIRCam observations. 
Following the theoretical predictions for the mass of these putative protoplanets, in Figure~\ref{fig: mass limit} we converted the NIRCam F410M contrast limits into mass limits using the \cite{Spiegel2012} evolutionary models for young planets with masses between $1$ and $4~M_{\rm Jup}$. These models also account for varying initial entropy \citep[hot-/warm-/cold-start models;][]{Spiegel2012}, different assumptions on the composition of the planetary atmosphere \citep[cloud-free spectra and hybrid-clouds spectra at solar metallicity;][]{Burrows2011}, and varying planetary ages, taken to be between 1 and 5 Myr for the HD~163296 system. 

The sky-blue and pink shades in Figure~\ref{fig: mass limit} illustrate the converted mass limits for planets inside the projected separation ranges of the D86 and D145 gaps, respectively (following the same shades as in Figure~\ref{fig: contrast limit}). 
The detection limit at a specific position depends on the position angle around the central star, and we investigated the empirical detection limit by injecting fake sources with varying magnitude based on the 5$\sigma$ contrast curves (Section~\ref{sec: Contrast Limits}). We focused on constraining the flux of the putative planets at the ALMA velocity kinks. At these locations, we derived the contrast limits of $\approx 8.3\times10^{-7}$ ($\Delta$F410M\ =\ 15.2~mag) for kink~\#1 and $\approx 3.3\times10^{-5}$ ($\Delta$F410M\ =\ 11.2~mag) for \#2, respectively (see Appendix~\ref{sec: Fake-source Injection at the Velocity Kinks}). The derived detection limit at kink~\#2 is different from the numerical contrast curve at the same separation ($6.9\times10^{-5}$ at $0\farcs67$) by a factor of $\sim2$ due to the presence of the R1 ring at similar separations. 
The derived F410M contrast limits correspond to a $\sim1-2\ M_{\rm Jup}$ hot-/warm-start gas giant at $\sim$5~Myr and a $4\ M_{\rm Jup}$ hot-start object at $\sim$1~Myr at kink \#1 and \#2, respectively.

Note that these estimates were obtained by assuming $A_{\rm V} = 0$, which is the estimated extinction for the star. As for the case of the source candidate discussed in the previous section, if the light of these putative planets is affected by significant extinction caused by the disk, the upper limit would be much higher. For example, the models presented by \citet[][]{Alarcon2024} provided estimates of $A_{\rm F410M} \sim 13$~mag for the putative planet at the kink~\#1 (private communication) and of $\sim39$~mag for the other planet at kink \#2. 
However, these models do not account for any perturbation on the disk material at the kink \#1 location, and the planet considered for their modeling at the kink \#2 has a very low mass of $0.1~M_{\rm{Jup}}$. If these planets are more massive, their interaction with the disk would likely remove a significant fraction of the disk material close to their orbits, and that would strongly decrease the estimated extinction. In fact, significantly lower values for the extinction can be obtained from the optical depths derived at $\lambda \approx 1$ mm from the analysis of the ALMA observations presented in \cite{Isella2016,Isella2018} \citep[cf.][]{Guidi2022}.
However, converting those optical depths into an extinction in the NIR/mid-IR requires an extrapolation over almost two orders of magnitude in wavelength and would be highly uncertain.
Future modeling work that incorporates these embedded planets is needed to derive more accurate extinction estimates at those specific locations.

\begin{figure*}
\centering
\begin{tabular}{cc}
    \begin{minipage}{0.5\textwidth}
        \centering
        \includegraphics[width=0.9\linewidth]{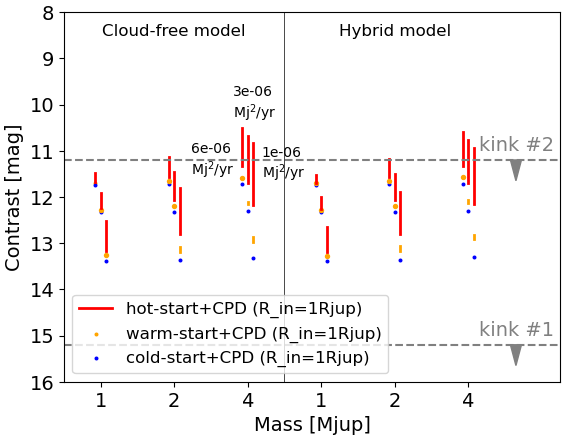}
    \end{minipage}%
    \begin{minipage}{0.5\textwidth}
        \centering
        \includegraphics[width=0.9\linewidth]{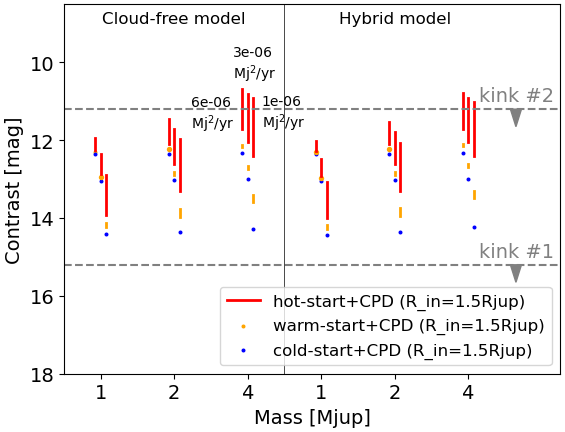}
    \end{minipage} \\
    \begin{minipage}{0.5\textwidth}
        \centering
        \includegraphics[width=0.9\linewidth]{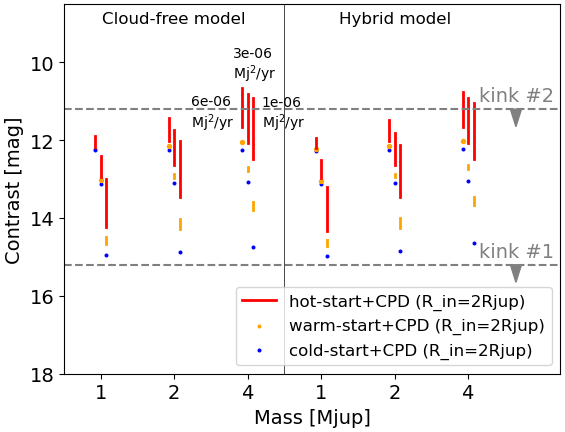}
    \end{minipage}
    \begin{minipage}{0.5\textwidth}
        \centering
        \includegraphics[width=0.9\linewidth]{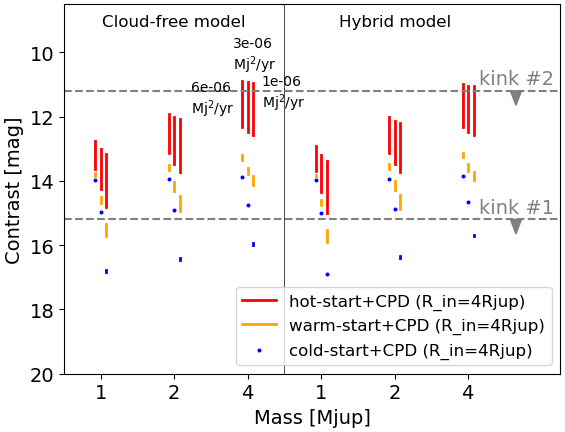}
    \end{minipage}
\end{tabular}
\caption{Same as in Figure \ref{fig: mass limit} with the addition of accreting CPD models \citep{Zhu2015}, calculated with different values for the disk inner radius ($R_{\rm in}=1, 1.5, 2, 4\ R_{\rm Jup}$, from top left to bottom right panels, respectively), and for the product of planet mass and mass accretion rate ($M_{\rm{p}} \dot{M}= 1, 3, 6\times10^{-6}\ M_{\rm Jup}^2/{\rm yr}$; for the order, see the annotations next to the data points of the $4\ M_{\rm Jup}$ cloud-free model).}
\label{fig: mass limit with CPD wit no extinction}
\end{figure*}

\subsubsection{Incorporating Circumplanetary Disk Models} \label{Circumplanetary}
We took into account possible thermal emission from a CPD around the putative young planets. We added the emission predicted by the \citet[]{Zhu2015} accreting CPD models to the hot-/warm-/cold-start models \citep{Spiegel2012} presented in the previous section.  
The comparison between these planet$+$CPD model predictions and our detection limits derived at the location of the two ALMA velocity kinks is shown in Figure~\ref{fig: mass limit with CPD wit no extinction} (the planet models are the same as in Figure~\ref{fig: mass limit}).
For the CPD models, we considered ranges for the product between planet mass and mass accretion rate ($M_{\rm{p}}\dot{M}=1-6\times10^{-6}\ M_{\rm Jup}^2/{\rm yr}$ assuming a planet forming within the stellar age), for the CPD inner radius ($R_{\rm{in}} = 1 - 4~R_{\rm{Jup}}$), and a CPD disk inclination of $i=47^\circ$, which assumes coplanarity with the protoplanetary disk.

Except for the case of a CPD with a relatively large inner radius $R_{\rm in}=4\ R_{\rm Jup}$, the F410M detection limit at kink~\#1 rules out a CPD model around a planet with mass $\geq 1\ M_{\rm Jup}$. A lower mass accretion rate than the values considered here would be required to have a warm- or cold-start $1\ M_{\rm Jup}$ planet surrounded by a CPD with a total flux consistent with our F410M detection limit.

\section{Conclusion} \label{sec: Conclusion}

We carried out JWST/NIRCam coronagraphic observations of the HD~163296 system with the F410M and F200W filters to search for infrared emission from the putative young planets predicted by models of the disk$+$planet interaction.
We employed roll-subtraction ADI reduction at two roll positions and utilized the spaceKLIP pipeline to remove stellar PSF.
After deliberate calibrations and post-processing, we did not detect robust signals of embedded protoplanets at the locations of the disk substructures observed with ALMA. We detected numerous background sources within the FoV, and several arc features that are likely tracing the forward-scattering side of more extended rings in the disk (R0--R3). Especially in the F200W we found an evident difference in the number density of background sources between regions within and beyond R3, which indicates significant extinction of the light from background sources by dust in the protoplanetary disk.

We also found a point-like source candidate at a separation $\rho\approx0\farcs75$ from the central star that is detected only in the F410M band at an SNR $\approx 4-5$ at KL=1--4 
with a contrast of $\Delta$F410M=11.8~mag. This corresponds to an apparent magnitude in the F410M band of 16.42$\pm0.15$~mag and to a lower limit for the F200W-F410M color of 5.5 referring to the previous VLT/SPHERE $K$-band observations. The measured flux is consistent with the predicted flux of models of young planets with a mass $\approx 2-4~M_{\rm{Jup}}$ without significant extinction from the disk material. 

We estimated contrast limits using spaceKLIP modules to derive upper limits for the mass of the putative young planets. Our F410M data achieved unprecedented azimuthally-averaged contrast limits at $\sim4\micron$ at $\rho\gtrsim0\farcs8$. In particular, we compared the NIRCam detection limits at the locations of the ALMA velocity kinks, where $\sim1-2 M_{\rm Jup}$ planets are predicted, with the predictions from evolutionary models for planets with different initial entropies \citep[hot-/warm-/cold-start models,][]{Spiegel2012}, with or without a circumplanetary disk. We found that the non-detection at kink \#1 requires 1) extinction by the disk and/or 2) cold or warm-start formation with a moderate accretion rate ($M_{\rm p} \dot{M} \lesssim 1\times10^{-6} M_{\rm Jup}^2/{\rm yr}$). Less stringent constraints were obtained on the mass of the putative planet causing the ALMA velocity kink \#2.

Follow-up observations and detailed modeling of the HD 163296 system are essential to confirm the nature of the point-like source candidate, and establish its membership to the system, as well as to investigate the disk morphology, extinction, and planet-disk interaction in detail.
Significant improvements are expected in the contrast limits that can be achieved with JWST in the future through RDI, including efforts to make empirical corrections on synthesized reference PSFs \citep{Lawson2024}, when reference stars will be available in the same configuration as our data.

Finally, we caution future direct imaging plans aiming at embedded protoplanets to acknowledge the risk of the potential extinction due to the protoplanetary disk. Observing (face-on) disks with a large cavity (e.g., PDS~70) or utilizing longer wavelengths, for instance, JWST/MIRI (Cugno et al. in prep) and future instruments/concepts such as Mid-infrared ELT Imager and Spectrograph \citep[METIS;][]{Brandl2010} mounted on the European Extremely Large Telescope (E-ELT), Mid-Infrared Camera, High-disperser, and IFU \citep[MICHI;][]{Packham2012} or Planetary System Imager \citep[PSI;][]{Fitzgerald2019} on the Thirty Meter Telescope (TMT), Thermal Infrared imager for the GMT which provides Extreme contrast and Resolution \citep[TIGER;][]{Hinz2012} on the Giant Magellan Telescope (GMT), Planet Formation Imager \citep[PFI;][]{Monnier2014}, and Large Interferometer For Exoplanets \citep[LIFE;][]{Quanz2022} could lower the risk of non-detections or insufficient sensitivities due to the disk extinction. Searching for CPDs at millimeter wavelengths would also be useful to advance the understanding of planet formation mechanisms \cite[e.g., with the Next Generation Very Large Array;][]{Ricci2018,Zhu2018}.


\section*{Acknowledgments}
The authors would like to thank the anonymous referee for their constructive comments and suggestions that improved the quality of the paper.
We thank Vanessa Bailey and Cornelis Dullemond for contributing to the JWST proposal (ID: GO~2540, PI: Luca Ricci).
We are grateful to Jorge Llop-Sayson for helping with synthetic RDI reduction, William Balmer for useful comments on the spaceKLIP steps, Felipe Alarcon for providing model extinction maps of the HD~163296 disk with the NIRCam filters, Gabriele Cugno for comprehensive discussions about JWST data, and Maxwell Millar-Blanchaer for his comments on the disk forward modeling.

This work is based on observations made with the NASA/ESA/CSA James Webb Space Telescope. The data were obtained from the Mikulski Archive for Space Telescopes at the Space Telescope Science Institute, which is operated by the Association of Universities for Research in Astronomy, Inc., under NASA contract NAS 5-03127 for JWST. These observations are associated with program GO 2540. The specific observations analyzed in this work can be accessed via \dataset[https://doi.org/10.17909/j87q-jf82]{https://doi.org/10.17909/j87q-jf82}.
Part of data presented here were obtained at the W. M. Keck Observatory, which is operated as a scientific partnership among the California Institute of Technology, the University of California and the National Aeronautics and Space Administration. The Observatory was made possible by the generous financial support of the W. M. Keck Foundation. We wish to acknowledge the critical importance of the current and recent Mauna Kea Observatory daycrew, technicians, telescope operators, computer support, and office staff employees, especially during the challenging times presented by the COVID-19 pandemic. Their expertise, ingenuity, and dedication are indispensable to the continued successful operation of these observatories.
The authors wish to recognize and acknowledge the very significant cultural role and reverence that the summit of Maunakea has always had within the indigenous Hawaiian community.  We are most fortunate to have the opportunity to conduct observations from this mountain.
This work has made use of data from the European Space Agency (ESA) mission
{\it Gaia} (\url{https://www.cosmos.esa.int/gaia}), processed by the {\it Gaia}
Data Processing and Analysis Consortium (DPAC,
\url{https://www.cosmos.esa.int/web/gaia/dpac/consortium}). Funding for the DPAC
has been provided by national institutions, in particular the institutions
participating in the {\it Gaia} Multilateral Agreement. This publication makes use of VOSA, developed under the Spanish Virtual Observatory (\url{https://svo.cab.inta-csic.es}) project funded by MCIN/AEI/10.13039/501100011033/ through grant PID2020-112949GB-I00.
VOSA has been partially updated by using funding from the European Union's Horizon 2020 Research and Innovation Programme, under Grant Agreement nº 776403 (EXOPLANETS-A).

Support for program GO~2540 was provided by NASA through a grant from the Space Telescope Science Institute, which is operated by the Association of Universities for Research in Astronomy, Inc., under NASA contract NAS 5-03127. Part of this research was carried out at the Jet Propulsion Laboratory, California Institute of Technology, under a contract with the National Aeronautics and Space Administration (80NM0018D0004). A.I. acknowledges support from the National Aeronautics and Space Administration under grant No. 80NSSC18K0828. S.Z. acknowledges support by NASA through the NASA Hubble Fellowship grant \#HST-HF2-51568 awarded by the Space Telescope Science Institute, which is operated by the Association of Universities for Research in Astronomy, Inc., for NASA, under contract NAS5-26555.


\appendix
\section{Comparison of the two masks with the F410M filter} \label{sec: Comparison of the two masks at F410M}
\begin{figure*}[h]
    \centering
    \includegraphics[width=0.9\textwidth]{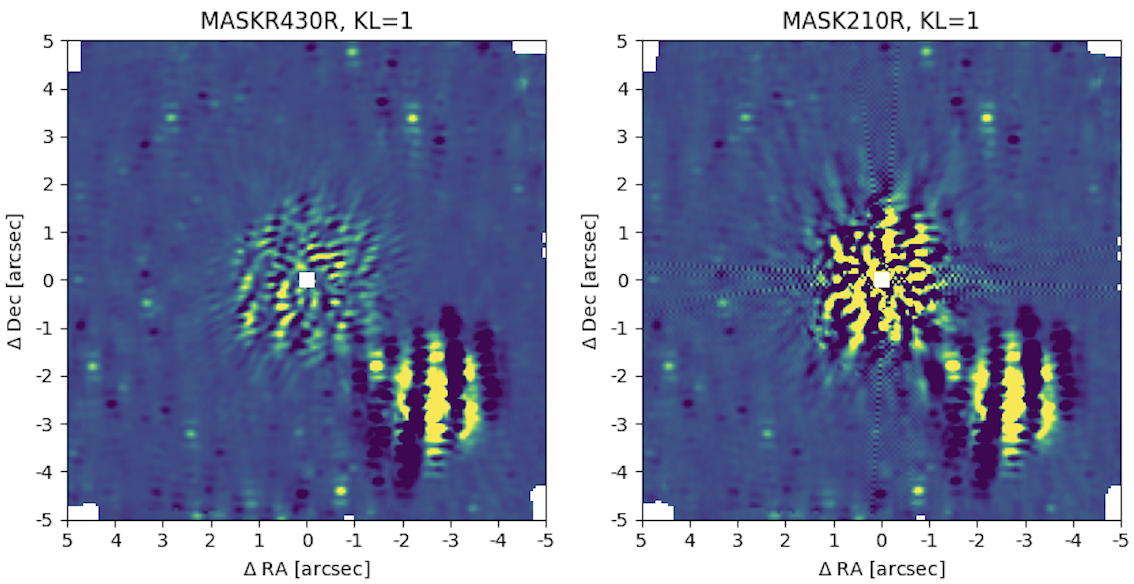}
    \caption{Comparison of the post-processed images of the primary mask (left) and the by-product  data for the F410M filter (right) in the same color scale.}
    \label{fig: F410M two masks}
\end{figure*}

Figure \ref{fig: F410M two masks} compares the post-processed images obtained from the primary data with F410M/MASK430R and the by-product data with F410M/MASK210R, as described in Section~\ref{sec: Data}. The F410M/MASK430R image was taken with the larger mask, and it suppresses residual speckles better than the other by-product data. Note that the PSF core of the central star in the MASK210R data set is saturated (showing NaN in the data), and spaceKLIP replaces the NaN value with 0 before image registration, which induced artifact features (marginal cross ripples). We reduced both data sets in the same manner and did not treat saturated pixels in the MASK210R data.

\section{Post-processing and forward modeling of the F410M data at different KL modes} \label{sec: Gallery of the post-processed images with different pyKLIP parameters}

\begin{figure*}[h]
    \centering
    \includegraphics[width=\textwidth]{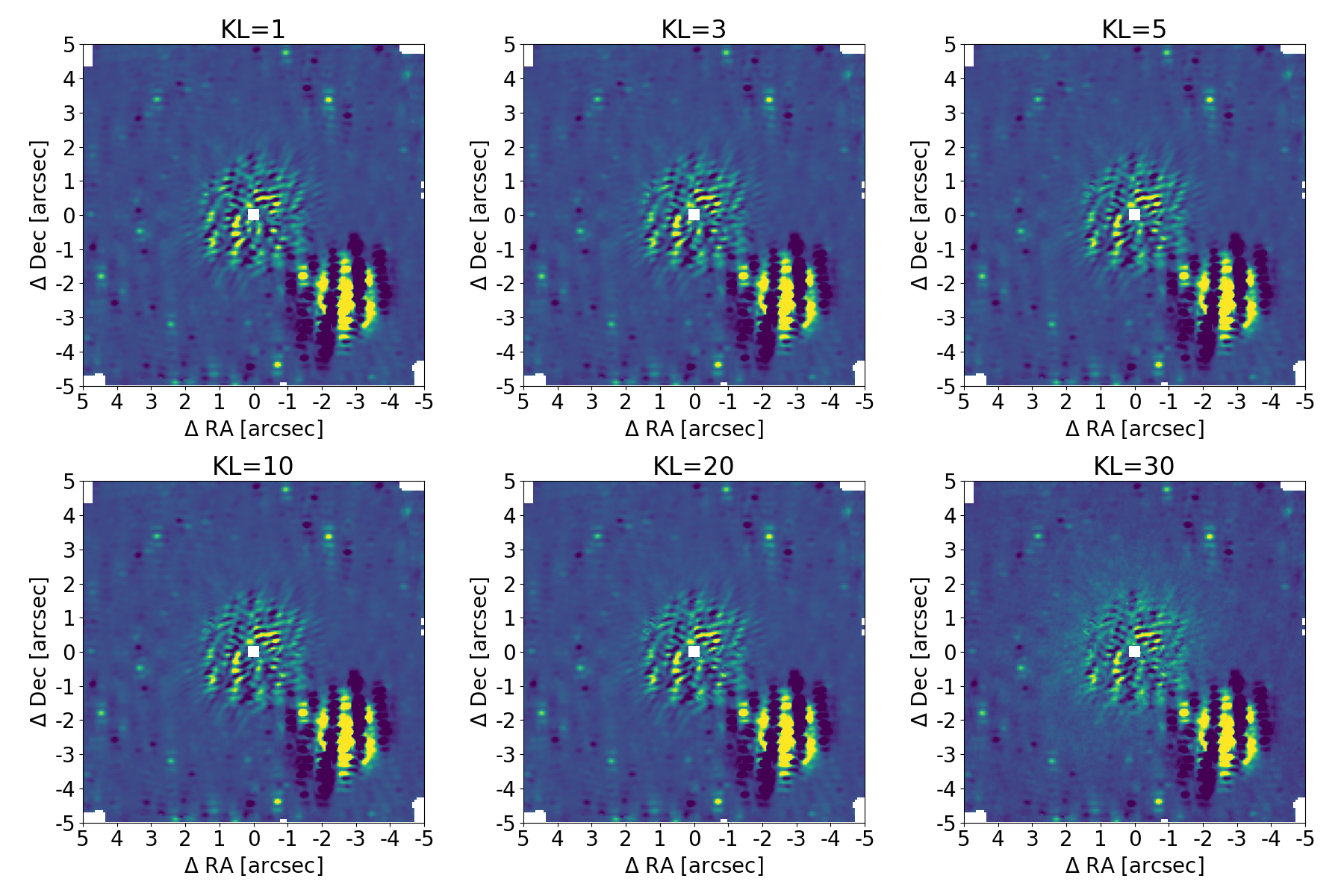}
    \caption{Gallery of the ADI post-processed images obtained with different KL modes, as labeled above each image.}
    \label{fig: gallery ADI}
\end{figure*}

Figure~\ref{fig: gallery ADI} shows the results of our ADI post-processing tests using KL values between 1 and 30. We found that more aggressive strategies for PSF subtraction with KL $>$ 10 do not greatly change the outcome of the post-processing images.

\begin{figure*}
    \begin{subfigure}
    \centering
        \includegraphics[width=\textwidth]{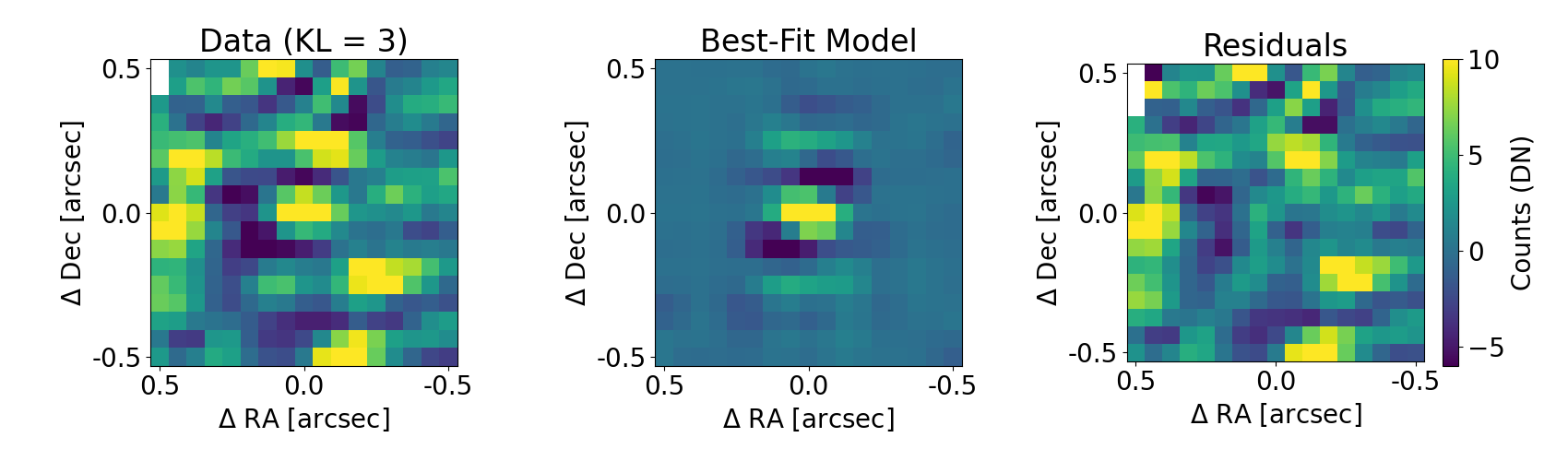}
    \end{subfigure}
    \begin{subfigure}
    \centering
        \includegraphics[width=\textwidth]{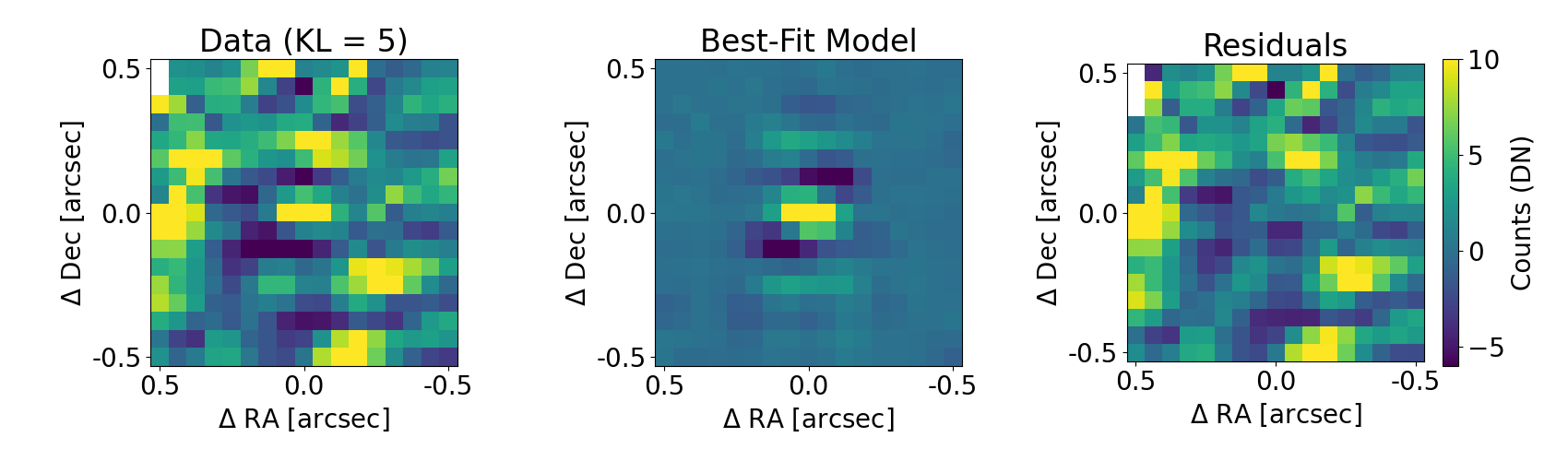}
    \end{subfigure}
    \begin{subfigure}
    \centering
        \includegraphics[width=\textwidth]{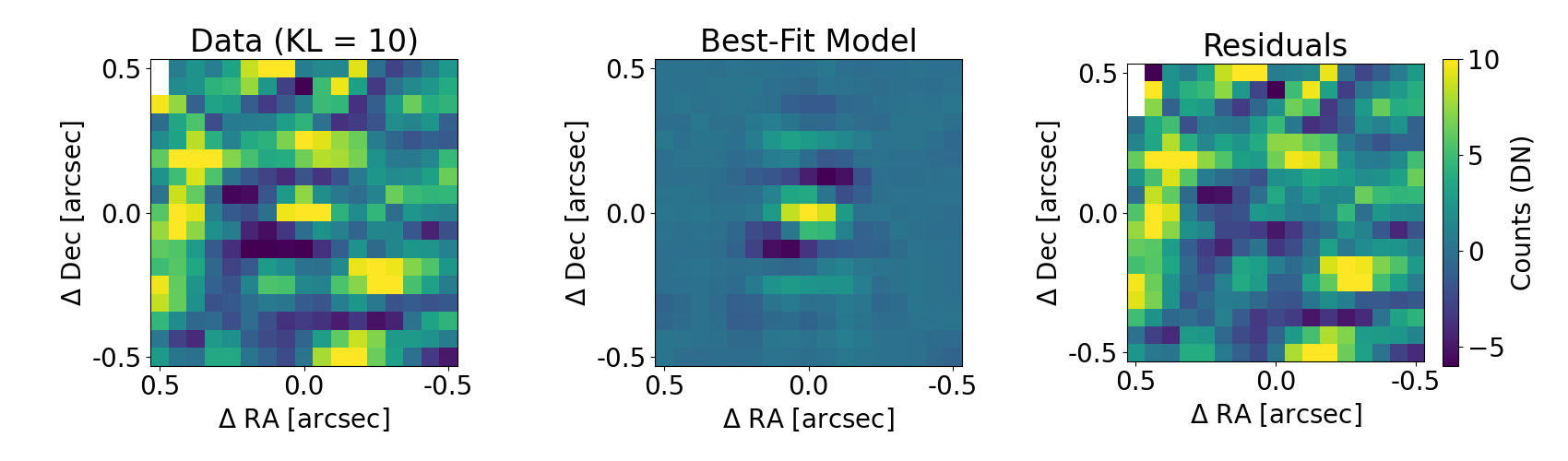}
    \end{subfigure}
    \caption{PSF fitting results (see Figure~\ref{fig: PSF fitting}) at KL\ =\ 3, 5, and 10 (from top to bottom).}
    \label{fig: gallery PSF fitting}
\end{figure*}

Figure~\ref{fig: gallery PSF fitting} shows some examples of PSF fitting at different KL modes. As mentioned in Section~\ref{sec: Point-like Source}, aggressive PSF subtraction at ${\rm KL}\geq5$ attenuates the signal of the source candidate as the angle offset between the two roll states is just $\sim10^\circ$.

\section{Attempt at RDI with a synthesized WebbPSF} \label{sec: Attempt at RDI with a synthesized WebbPSF}

\begin{figure}[h]
    \centering
    \includegraphics[width=0.9\textwidth]{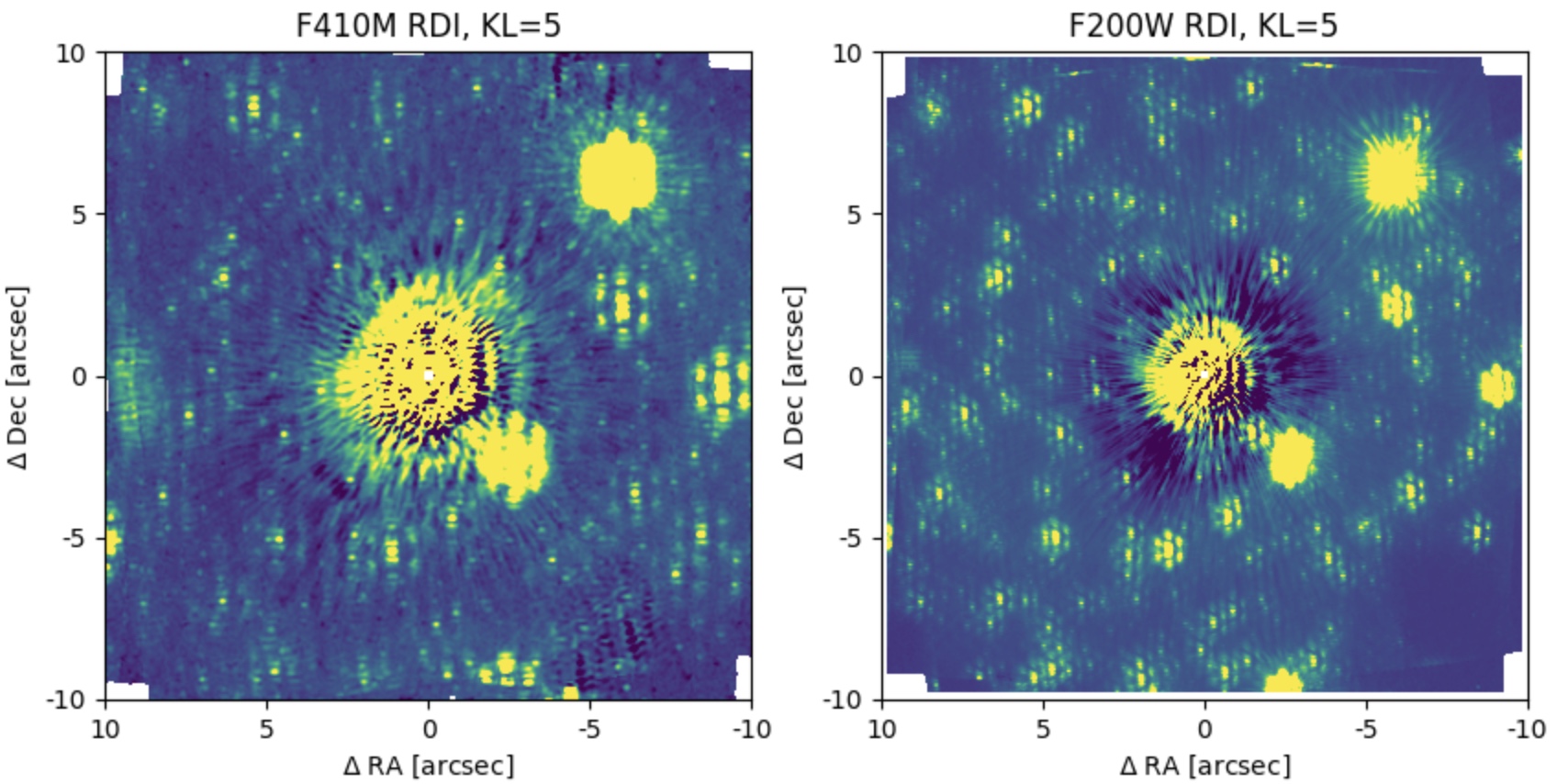}
    \caption{RDI post-processed images in the F410M filter (left) and F200W filter (right) after generating synthesized reference PSFs from WebbPSF, and incorporating the wavefront errors close to the observing epoch and the observational configurations. Note that we conducted RDI with the full FoV in each filter. The inner regions at separations smaller than 3\farcs5 are dominated by residual speckles and artifacts by the modeled reference PSF.}
    \label{fig: test RDI}
\end{figure}

For JWST coronagraphic imaging, RDI is generally more efficient than ADI in detecting companions \citep{Carter2023,Hinkley2023}, but it is hard to find a suitable reference star that has a similar color to a young star with a disk. 
Our program did not observe a reference star, and we attempted synthetic RDI by generating reference PSFs via WebbPSF \citep{Greenbaum2023}. We incorporated the wavefront errors measured on the closest date of our observations and with the same instrumental configurations. However, the synthetic PSFs we generated in this way proved to be not optimal for RDI post-processing as shown in Figure~\ref{fig: test RDI}, where strong speckle residuals and several over-subtraction features are evident in both filters. 

\section{Forward modeling of the disk features} \label{sec: Forward modeling of the disk features}

\begin{figure}[h]
    \centering
    \includegraphics[width=0.9\textwidth]{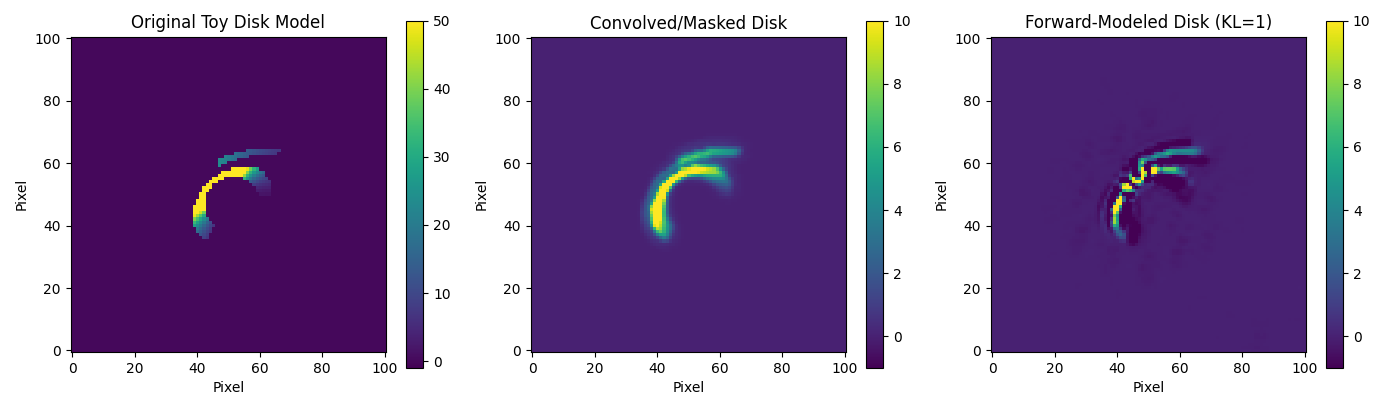}
    \caption{Comparison between the image of the toy disk model presented in Appendix~\ref{sec: Forward modeling of the disk features} (left), the model image after convolution using WebbPSF at F410M and MASK430R throughput (middle) and a forward-modeled disk, which simulates the post-processing methods applied to our JWST/NIRCam observations in the F410M filter (right, KL=5). The forward-modeled image suggests that the bright disk feature convolved with WebbPSF can introduce point-like diffraction patterns at different locations from the disk feature (e.g., bright pixels at the lower right in the right panel) but does not introduce significant additional extended features.}
    \label{fig: fwdmd disk}
\end{figure}

\begin{figure}[h]
    \centering
    \includegraphics[width=\textwidth]{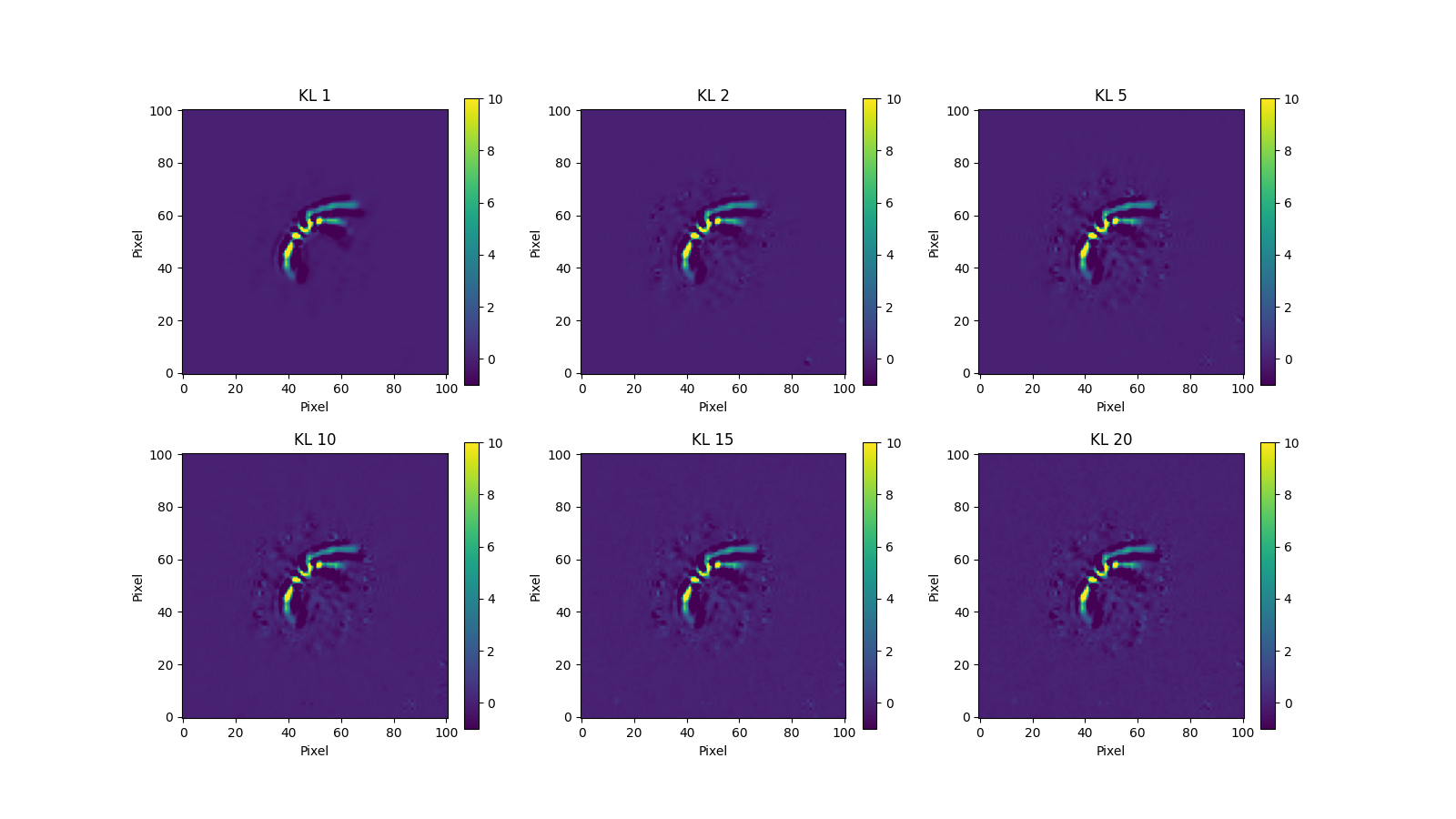}
    \caption{Gallery of the forward-modeled disk (Figure~\ref{fig: fwdmd disk}) with different KL modes.}
    \label{fig: gallery fwdmd disk}
\end{figure}

As described in Section~\ref{sec: Disk Features} we tested forward modeling using a toy disk model to interpret the F410M post-processed image. 
We refer to \cite{Rich2019} for a model for the R1 ring, and slightly modified the inner and outer edges and the offset from the star to better match the arc-like feature seen in the F410M image. A second partial ring was added to the model to reproduce the R2 feature, as well as to test if additional arc-like features in the F410M map can be reproduced. Before running forward-modeling modules, we convolved the toy model with an off-axis WebbPSF, in the same way as in Appendix~\ref{sec: Attempt at RDI with a synthesized WebbPSF}, and applied the mask throughput map, which was modeled with the spaceKLIP processes. Note that the actual NIRCam PSF shape near the coronagraph mask is a function of separation from the mask center, and detailed disk modeling requires convolving the model with on-axis PSFs \citep[e.g.,][]{webbpsf_ext}.
We finally used the Disk Forward Modeling modules in {\tt pyklip} \citep[DiskFM;][]{DiskFM} for forward modeling.
Figure~\ref{fig: fwdmd disk} presents our toy model, a WebbPSF-convolved and mask throughput-applied disk image, and the resulting map for the forward-modeled disk, which simulates the post-processing methods applied to our JWST/NIRCam observations in the F410M filter. Figure~\ref{fig: gallery fwdmd disk} is a gallery of the forward-modeled disk at different KL modes. We found that these bright disk features introduce some local excesses at different locations from the rings (e.g., some relatively bright pixels in the south-west side from the star), but they are not stable among different KLs. Furthermore, we did not detect additional extended features than what we modeled, suggesting that the R2 feature is not induced by diffraction patterns of produced by the R1 ring.

\section{Keck/NIRC2 $M_s$-band Observations} \label{sec: NIRC2}

JWST and ground-based coronagraphic observations play complementary roles for the search for companions. At inner separations (extreme) adaptive optics combined with optimized coronagraphs have achieved higher contrast than JWST coronagraphy \citep[e.g.,][]{Ren2023} while JWST can achieve very high sensitivity at larger separations.

\begin{figure*}
\centering
\begin{tabular}{cc}
\begin{minipage}{0.5\hsize}
    \centering
    \includegraphics[width=0.9\textwidth]{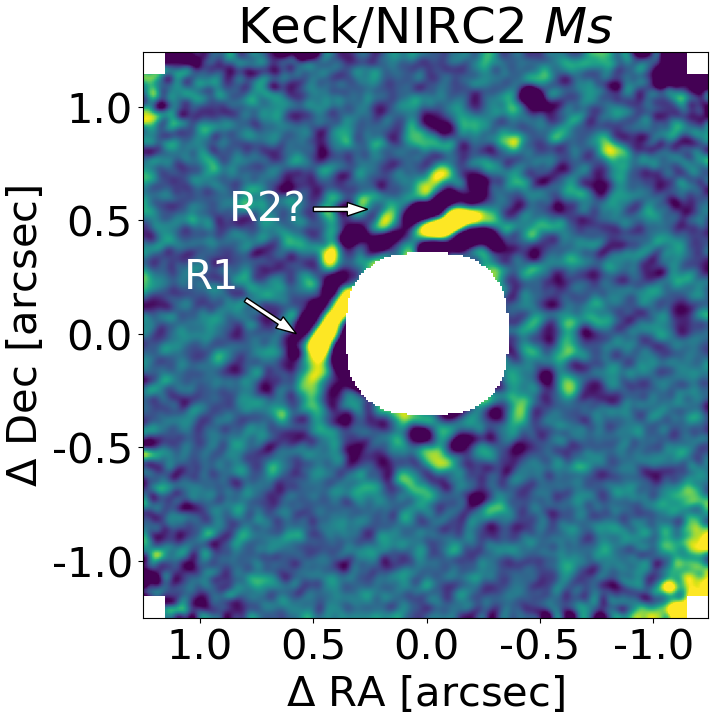}
\end{minipage}
\begin{minipage}{0.5\hsize}
    \centering
    \includegraphics[width=0.9\textwidth]{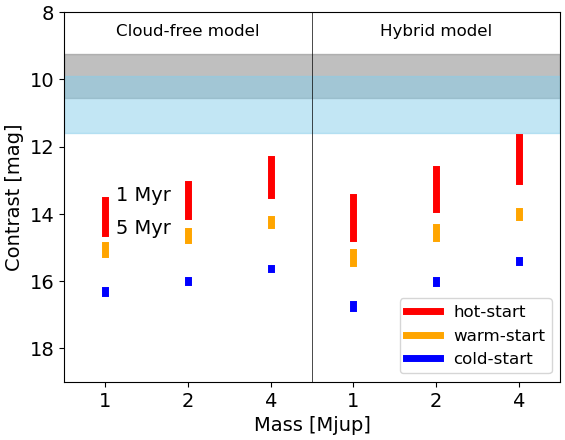}
\end{minipage}
\end{tabular}
\caption{Left) Keck/NIRC2 $Ms$-band post-processed image after smoothing with a Gaussian with $\sigma=2$~pix to better highlight the extended features. Right) Diagram showing the conversion between the NIRC2 $Ms$-band contrast limits and the mass from evolutionary models of planets with different properties (as in Figure~\ref{fig: mass limit}; the colored shades indicate the contrast limits at the projected separations of the ALMA dust gaps (gray for D10 and sky-blue for D45); the colored bars represent the hot- (red)/warm- (orange)/cold-start (blue) models \citep{Spiegel2012} with cloud-free and hybrid clouds atmospheres at solar metallicity \citep[][]{Burrows2011}, and ages between 1 and 5~Myr (the upper and lower limits of each model range correspond to 1 and 5~Myr, respectively), assuming the same extinction as the central star, i.e. $A_{\rm V}=0$.}
\label{fig: NIRC2 results}
\end{figure*}

In this study we reduced archival Keck/NIRC2 $Ms$-band (pivot wavelength: $\lambda=4.670\ \micron$, bandwidth:$\Delta\lambda=0.241\ \micron$) data (PI: Garreth Ruane, date: 2020 August 10 UT, total exposure time: 7500~seconds). 
The data were taken with the vortex coronagraph \citep{Serabyn2017} and vertical angle mode for ADI. After calibrating the raw data (flat fielding, distortion correction, sky subtraction, image alignment), we cropped the image by $\sim2\farcs5\times2\farcs5$ focusing on inner separations, where Keck/NIRC2 can outperform JWST/NIRCam high-contrast imaging (see also Section~\ref{sec: Contrast Limits} for the practical contrast limits), and then utilized {\tt pyklip} post-processing algorithms to perform ADI reduction. The left panel in Figure~\ref{fig: NIRC2 results} shows the post-processed NIRC2 image with a clear detection of R1 and a marginal detection of R2.

As described in Section~\ref{sec: Contrast Limits}, at separations $\rho \lesssim0\farcs8$ covering the ALMA dust gaps D45 and D86, the Keck/NIRC2 observations achieved a better contrast level than the NIRCam/F410M observations.
However, the NIRC2 contrast limits did not achieve enough contrast to be sensitive to planets with masses $\lesssim~5~M_{\rm{Jup}}$ at the D45 and D86 locations (see the right panel in Figure~\ref{fig: NIRC2 results}). In contrast, the F410M data achieved better mass limits than the NIRC2 data in the D86 gap as well as at larger separations (see Section~\ref{sec: Constraining the Mass of the ALMA Putative Planets}).

We also note that at the location of the source candidate discussed in Section~\ref{sec: Point-like Source}, the NIRC2 data did not reach enough contrast to detect this source. We injected fake sources assuming the same contrast level to the F410M data ($1.83\times10^{-5}$) at the same separation (0\farcs754) and at four position angles in the data set and reran the {\tt pyklip} reduction, but did not reproduce the injected sources.

\section{ALMA CO velocity maps} \label{sec: Companion candidate vs ALMA CO velocity maps}

We downloaded the ALMA $^{12}$CO ($J=2-1$) molecular gas emission data of HD~163296 from the MAPS webpage\footnote{\url{https://alma-maps.info/data.html}}, which has a higher velocity resolution than the DSHARP CO data \citep{Isella2018}.
Figure \ref{fig: F410M vs ALMA CO, MAPS} shows the location of the source candidate detected in the NIRCam/F410M data (see Section~\ref{sec: Point-like Source}) on the ALMA channel maps. Figure~\ref{fig: radial profile of CO, MAPS} in the main text shows a comparisons of the radial profiles of the CO gas emissions with the separation of the source. As mentioned in \cite{Izquierdo2023}, the $^{12}$CO emissions have a dip at a similar separation to the source candidate.

\begin{figure*}
    \centering
    \includegraphics[width=0.9\textwidth]{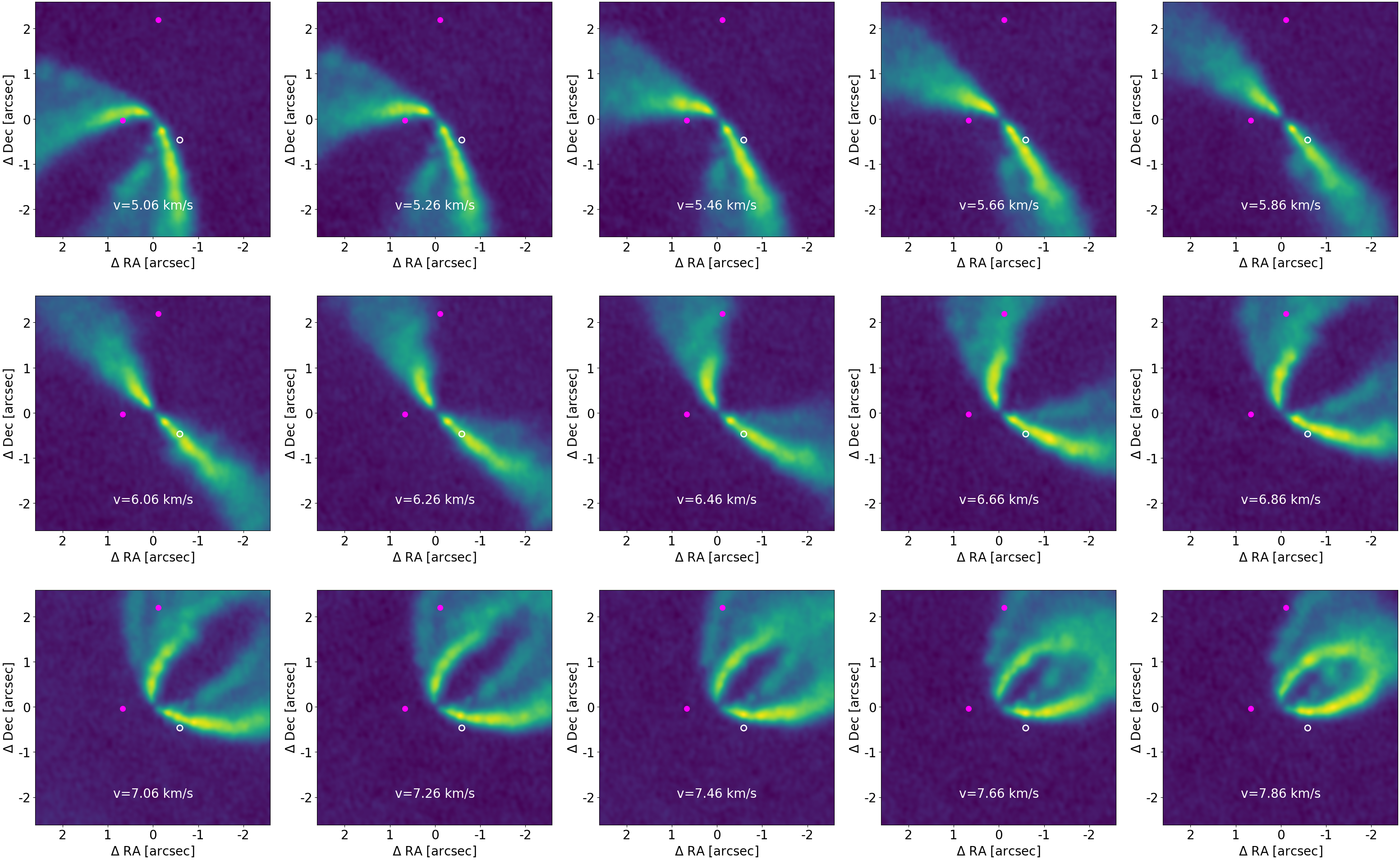}
    \caption{Comparison of the source candidate (white) as well as the velocity kinks (magenta) with the $^{12}$CO ($J=2-1$) channel maps obtained by MAPS \citep[FWHM of the beam size: $0\farcs15$;][]{Oberg2021}.}
    \label{fig: F410M vs ALMA CO, MAPS}
\end{figure*}

\section{Fake-source Injection at the Velocity Kinks} \label{sec: Fake-source Injection at the Velocity Kinks}

\begin{figure*}
    \centering
    \includegraphics[width=0.95\textwidth]{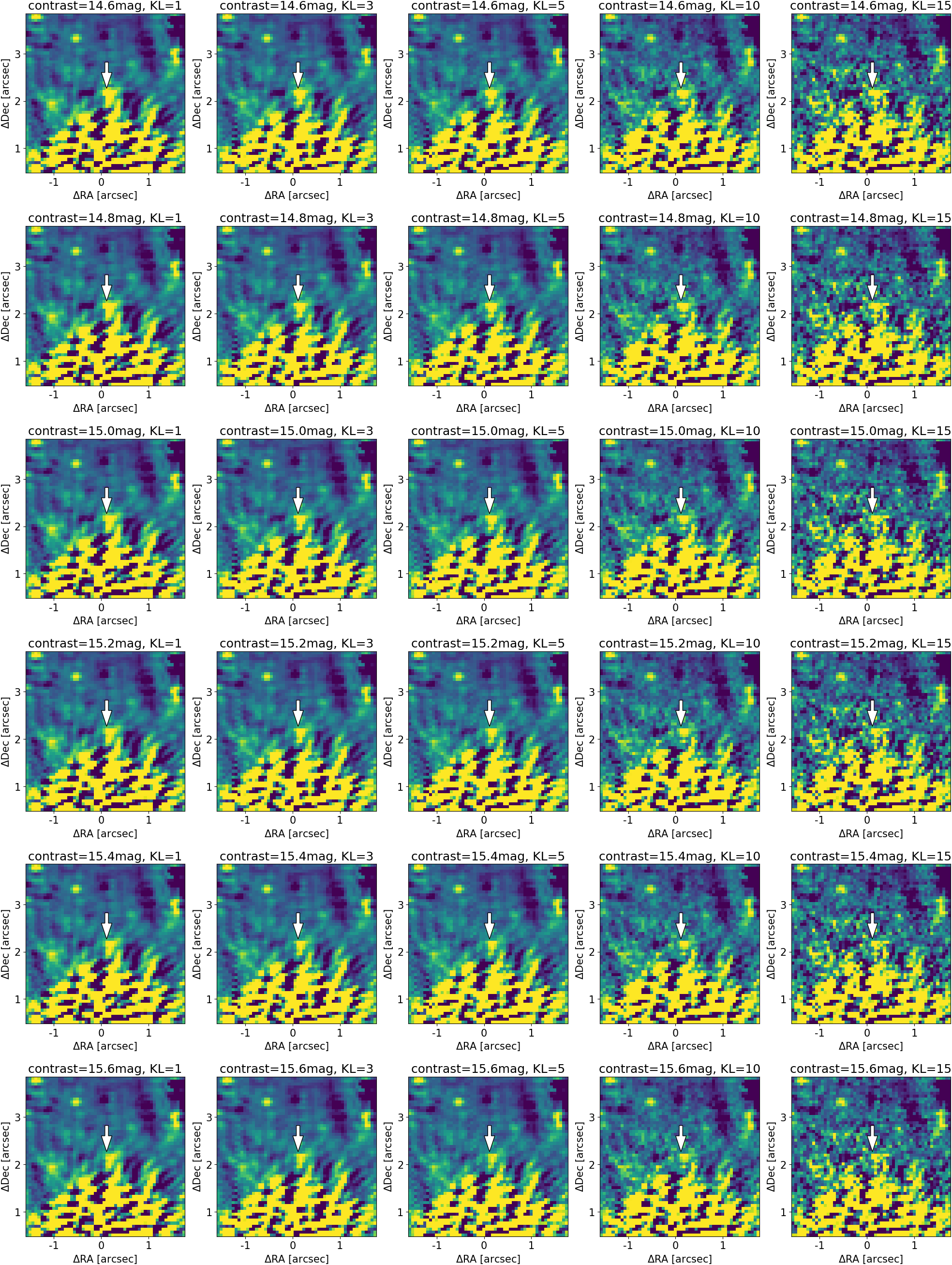}
    \caption{F410M post-processed images with different KL modes after injecting a fake source with different contrast levels ($\Delta$F410M\ =\ 14.6--15.6~mag) at the location of the ALMA velocity kink~\#1. The injected source is indicated by a white arrow in each image.}
\label{fig: fake-source injection kink 1}
\end{figure*}

\begin{figure*}
    \centering
    \includegraphics[width=0.95\textwidth]{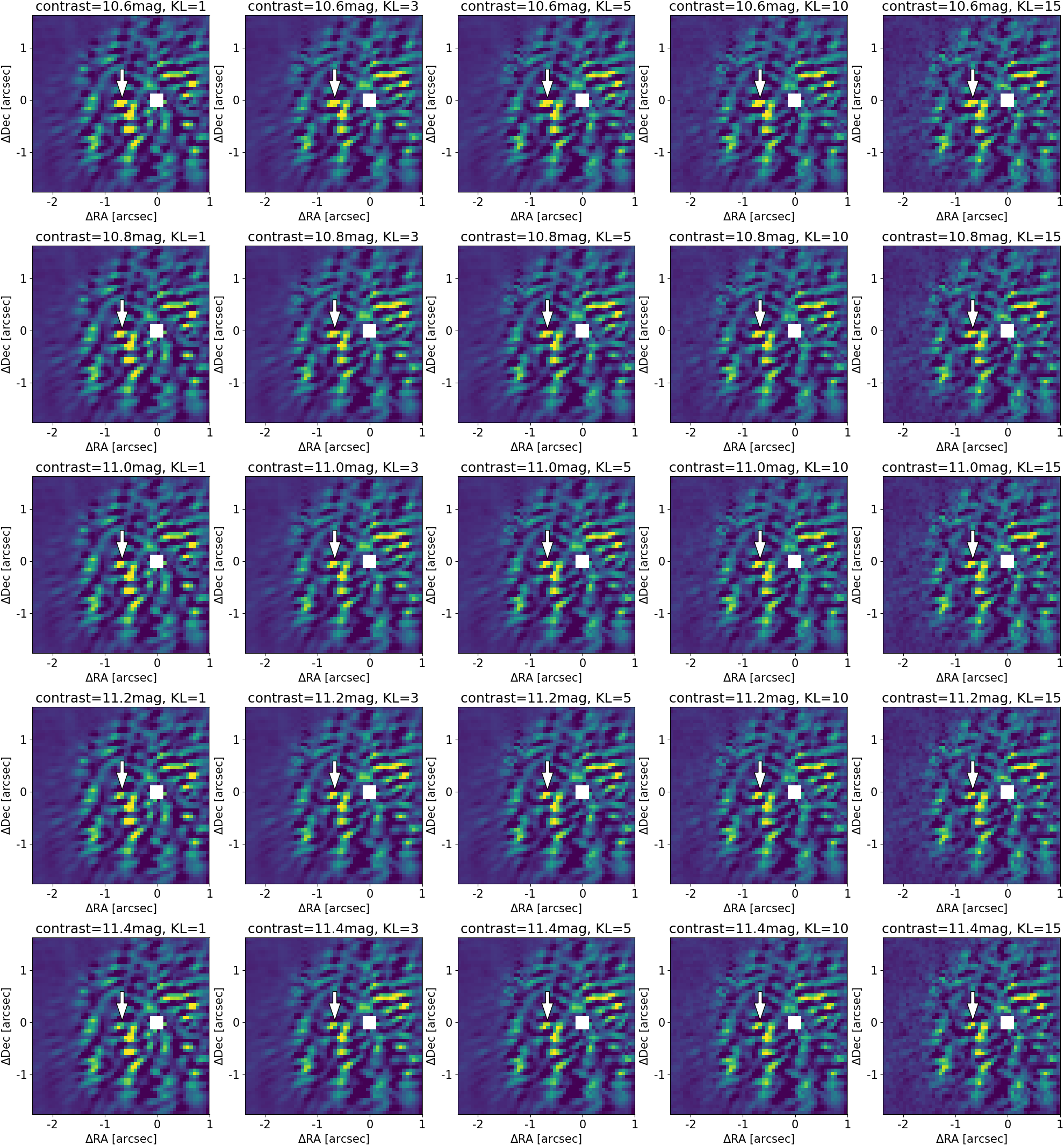}
    \caption{Same as in Figure~\ref{fig: fake-source injection kink 1} but with injected fake sources at the location of the ALMA velocity kink~\#2 (with contrast levels $\Delta$F410M\ =\ 10.6--11.4~mag).}
\label{fig: fake-source injection kink 2}
\end{figure*}

We utilized {\tt analysistools.extract\_companions} to inject positive fake sources at specific positions with the goal of estimating the empirical detection limits at the locations of the ALMA velocity kinks~\#1 and \#2.
Figures~\ref{fig: fake-source injection kink 1} and \ref{fig: fake-source injection kink 2} show the fake-source injection results at different contrast levels and KL modes. We adopted the empirical detection limits at the locations of the velocity kinks at the contrast levels at which we could spatially resolve the injected source from residual speckles and the disk features with KL\ =\ 5.

Due to the relatively small field rotation, aggressive strategies for PSF subtraction with KL$\geq10$ induce more residuals of stellar light and speckles at $\rho\lesssim2\arcsec$, while the injected fake source at kink~\#1 is well retrieved at smaller KL modes (see Figure~\ref{fig: fake-source injection kink 1}). For the fake-source injection at kink~\#2, the location is so close to R1 that the PSF has an overlap with it. However, the derived empirical detection limit is better than the numerical contrast limit, which in turn is affected by the presence of R1 (see Section~\ref{sec: Constraining the Mass of the ALMA Putative Planets}).

\bibliography{library}                                    
\end{document}